\documentclass[useAMS,usenatbib,usegraphicx]{mn2e}

\usepackage{subfigure} 
\usepackage{times}

\def\aaps{A\&AS}
\def\apj{ApJ}
\def\apjl{ApJL}
\def\mnras{MNRAS}
\def\pasa{PASA}
\def\pasp{PASP}
\def\aj{AJ}
\def\araa{ARA\&A}
\def\aap{A\&A}
\def\apjs{ApJS}
\def\nat{Nature}

\def\hi        {\hbox{H{\sc i} }}
\def\hie       {\hbox{H{\sc i}}}
\def\hipass    {\hbox{H{\sc i}PASS }}
\def\arcdeg    {\ensuremath{^{\circ}}}          
\def\msun      {\ensuremath{M_{\odot} \;}}      
\def\nhi       {\ensuremath{N_{H\!I}}}            

\def\kms       {\hbox{km~s$^{-1}$~}}            
\def\kmse      {\hbox{km~s$^{-1}$}}
\def\ni        {\noindent}

\title[Correlations among galaxy properties found in a blind \hi Survey]
{Correlations among the properties of galaxies found in a blind \hi Survey, which also have SDSS optical data.}

\author[D. A. Garcia-Appadoo et al.]
{D. A. Garcia-Appadoo$^{1,4}$\thanks{E-mail: dgarcia@eso.org}, A. A. West$^{2,3}$, J. J. Dalcanton$^{3}$, L. Cortese$^{4}$ 
\newauthor and M. J. Disney$^{4}$\\
$^{1}$European Southern Observatory, Alonso de Cordova 3107, Casilla 19001, Vitacura, Santiago 19, Chile\\
$^{2}$Astronomy Department, University of California, 601 Campbell Hall, Berkeley, CA 94720-3411, USA\\
$^{3}$Physics-Astronomy Building C309, Department of Astronomy,  University of Washington, Box 351580, Seattle, WA 98195, USA\\
$^{4}$Physics \& Astronomy Department, Cardiff University, 5 The Parade, Cardiff, CF24 3AA, UK}
\begin{document}

\date{Accepted. Received; in original form}

\maketitle

\label{firstpage}

\begin{abstract}

We have used the Parkes Multibeam system and the Sloan Digital Sky Survey (SDSS) to 
assemble a sample of 195 galaxies selected originally from their \hi signature to 
avoid biases against unevolved or low surface brightness objects. For each source 9 
intrinsic properties are measured homogeneously, as well as inclination and an 
optical spectrum. The sample, which should be almost entirely free of either 
misidentification or confusion, includes a wide diversity of galaxies ranging from 
inchoate, low surface brightness dwarfs to giant spirals. Despite this diversity 
there are 5 clear correlations among their properties. They include a common 
dynamical mass-to-light ratio within their optical radii, a
correlation between surface-brightness and Luminosity and a common 
\hi surface-density. Such correlation should provide strong constrains on models of 
galaxy formation and evolution.
\end{abstract}

\begin{keywords}
galaxies: fundamental parameters -- galaxies: evolution -- galaxies: general -- galaxies: structure -- galaxies: peculiar.
\end{keywords}

\section{Introduction}

Searching for systematic correlations among the global properties of galaxies, 
analogous to the H-R diagram for stars, may offer the best hope of
having a better understanding about their formation and evolution. The recent availability of large 
and systematic data-sets at several wavelengths makes this a good time to search. 
Such searches go back to the optical pioneers of galaxy exploration such as Hubble (1937), 
Holmberg (1965), de Vaucouleurs (1991) and Zwicky (1942). The trends and correlations they discovered, 
re-emerge very clearly in the Sloan Digital Sky Survey, as described by Blanton et al. (2003) 
in their analysis of 200.000 galaxies at redshifts of $\sim$ 0.1.

This paper deals with a much smaller sample of galaxies, but ones found by an 
entirely different technique, i.e. in a blind search for neutral Hydrogen gas at 
21-cm. The principal motivations for such a blind search are two. First of all, by 
definition, optically selected galaxies already contain many stars, which may not 
be the case in the younger, or less evolved galaxies, which may have the most to tell 
us about their formation and evolution. Apart from X-ray searches, which detect the 
very hot gas in the potential wells of giant ellipticals, most galaxy searches 
rely, either directly or indirectly, on the presence of stars and so are biased against young 
or unevolved objects. Secondly, a blind \hi search offers a unique way round the 
strong optical surface brightness selection effects, which could disguise the very correlations one 
is looking for. By definition a galaxy must be separately both luminous enough
\emph{and} large enough to distinguish it above a sky background which, by galaxy 
standards, is quite bright. These two \emph{separate} requirements squeeze optically selected 
galaxies into an extremely narrow range of surface brightnesses (SB) centred, for a 
given catalogue, on $\Sigma_{cat}$ where:
\begin{equation}
\Sigma_{cat} \equiv \frac{l_{ap}}{\pi\theta^{2}_{ap}}
\end{equation}
where $l_{ap}$ and $\theta_{ap}$ are the minimum apparent luminosity, and the minimum 
apparent angular radius which the galaxy catalogue will accept (Disney 1999). The Full Width Half Maximum 
(FWHM) of the detectable range is typically only 3 magnitudes wide. Such a narrow 
stricture certainly impoverished the old photographic surveys (Disney 1976, Disney \& 
Phillipps 1983, Dalcanton 1997). It might be naively thought that CCDs would have cured this 
stricture. Not so, because they detect galaxies that are both fainter \emph{and} 
smaller than emulsions, their $\Sigma_{cats}$'s are (see equation (1)) scarcely any dimmer. On the other 
hand, the limitation with an \hie-selected sample is that it will miss the early type galaxies - which 
contain very little neutral gas. According to the SDSS survey late types make up the large 
majority of all galaxies, although they may emit less than their fair share of light (Blanton et al. 2001).

Blind \hi surveys face four main problems. To reach sources of a given column density of \hi (N$_{H\!I}$ in 
atoms cm$^{-2}$) any survey must be at least sensitive enough to find them in the most favourable case, 
i.e. when they fill the beam. In that case (Minchin 2003, Disney 2008) t$_{obs}$ (per beam) $\ge$ 
$k$/[N$_{H\!I}$ (min)]$^{2}$ where $k$ is independent of dish-diameter D (because a larger dish projects 
the same system noise onto a smaller area of sky). Since there is a loose correspondence between N$_{H\!I}$ 
and SB (e.g. Swaters et al. 2003). Mathematically:
\begin{equation}
N_{H\!I} \cong 10^{20.1} \left(\frac{M_{H\!I}}{L_{B}}\right)_{\odot} 10^{[0.4(27-{\overline{\mu}}_{21})]}~atoms~cm^{-2}
\end{equation}
(Disney \& Banks 1997) where $\left(\frac{M_{H\!I}}{L_{B}}\right)_{\odot}$ is in solar units and 
$\overline{\mu}_{21}$ is the average SB in B mag arcsec$^{-2}$, taken over the same area as N$_{H\!I}$. Since 
the effective SB ($\mu_{ef\!f}$) of an exponential disc is 2.1 magnitudes brighter than $\overline{\mu}_{21}$ 
(Salpeter \& Hoffman 1996) column density sensitivity at the $N_{H\!I} < 10^{19.5} cm^{-2}$ level is 
needed to detect LSBGs (i.e. $\mu_{ef\!f}$ $\>$ 25.0 B mag arcsec$^{-2}$) with $\left(\frac{M_{H\!I}}{L_{B}}\right)_{\odot}$ 
$\approx$ 0.3 (see Section 3), which in turn requires integration times per beam of several hundred 
seconds - which rendered earlier blind \hi surveys either too insensitive or too impractical to 
detect LSBGs. Since this difficulty was not recognised until recently, some earlier claims, based on small 
blind \hi surveys, to set rigid upper limits to the amount of cosmic \hi or the numbers of LSBGs in the 
Universe, must be set aside [e.g. Shostak 1977, Fisher and Tully 1981, Zwaan et al. 1997]. 

The coming of the Multibeam \hi detector (Staveley-Smith et al. 1996) was essential 
to carry out such blind \hi surveys. HIPASS, the \hi Parkes All Sky Survey (Meyer et al. 2004) used the first 
multibeam system to survey the entire Southern Sky, and the north up to +25$^{\circ}$. More than 4000 
sources were identified in the Southern Sky alone. The Equatorial Survey, described in this paper, 
was initially part of HIPASS, but with an accelerated search so as to exploit the SDSS-DR2 optical data when 
it emerged. HIPASS sources are typically more than a degree apart which makes for a real challenge 
in obtaining complementary optical, and other data, simply because, before SDSS, each source required a 
separate observing campaign.

The strong clustering of galaxies, and of \hi galaxies in particular, makes the identification of the 
source with an optical candidate quite challenging. This is true even when both 21-cm and optical velocities 
are known, for galaxy velocities are strongly clustered too. To be certain that every Parkes-detected 
source was correctly identified with its optical counterpart would require interferometric follow-up in 
every case - which is infeasible when one is dealing with hundreds of sources, as here. We have used 
interferometry, coupled with simulations, to be certain that only a handful of the remaining sources (less than 10) are 
still misidentified. This handful is too small to affect the main correlations. 

The Equatorial Strip Survey (ES henceforth), reported on here, was a search through HIPASS cubes 
between declinations $-$6\arcdeg and +10\arcdeg. Thus 5780 deg$^{2}$ of \hi data, in the velocity range between 
$-$1280 and +12700 \kmse, were searched largely by eye to come up with 1107 sources (Garcia-Appadoo PhD Thesis 2005). The 
Equatorial Strip was chosen because: (i) it was approximately perpendicular to the Galactic Plane and so it is mostly dark 
enough to make LSBGs detectable; (ii) it is accessible for follow up with a large range of instruments including the VLA, 
and for some of its area the SDSS-DR2; (iii) it includes the area searched for LSBGs by Impey et al. (1996) using the 
APM machine (Cawson et al. 1987) to scan UK-Schmidt plates. This should eventually lead to an estimate of the number 
of LSBGs still missing from wide-scale optical surveys. So long as LSBGs remain a putative reservoir for missing baryons 
(e.g. Fukugita et al. 1998) it is important to know this.
Of the ES area, 50 per cent will eventually be scanned by SDSS, and 35 per cent of it is already publicly released. 
We concentrated our search in the 1700 deg$^{2}$ of the ES released earlier in DR2 (Abazajian et al. 2004). 370 cross 
identifications were made, and then refined down to only 195 to reduce possible misidentifications to a minimum.  These 
195 sources have SDSS-DR2 optical diameters (50 per cent light diameter and 90 per cent light diameter) between 0.16 and 
6 arcmins (West et al. 2008). The SDSS pipeline photometry on sources of this size was known to be extremely unreliable so 
we had to devise new techniques to reduce the data (West PhD Thesis 2005), which delayed the project by nearly two years. 
Eventually we will analyse many more ES sources already observed with SDSS. But the results found for the first 195 are 
clear and interesting enough to merit being published now. 
     
We have measured, at 21-cm, the peak flux, the integrated flux, the line width, and the spectral shape (e.g. 
two horned or Gaussian mainly), and in the optical the luminosities at $u, g, r, i, z,$ two 
radii R$_{50}$($g$) and R$_{90}$($g$) containing respectively 50 and 90 per cent of the $g$-band light, the inclination, 
the morphology and a fibre nuclear spectrum set either on the nucleus or the brightest HII region.  
Given the strong correlations between some of the colours this amounts to about a dozen 
independent measurements, not far short of the 14 desired. Chiefly missing is a rotation 
curve to spell out the distribution of Dark Matter. Nevertheless we have sufficient information 
to hope that some important aspects of galaxy systematics will emerge. With 13 properties there 
will be $\sim$ (13$\times$12)/2 $\sim$ 80 possible correlations to look for.

The previous largest survey of discs was reported in a remarkable, but largely unremarked 
paper by Gavazzi et al. in 1996 entitled \emph{The phenomenology of disc galaxies}. They found: 
(a) the mass-to-light ratio in the NIR is virtually constant and equal to 4.6 in H-band solar units;
(b) Population I indicators are all anti-correlated with mass; 
(c) Conversely the luminosity and surface brightness of old stellar populations all increase with mass; 
(d) Bulge components all increase non-linearly with mass; 
(d) All the above properties are independent of either morphological type or environment.   

Previous blind \hi surveys, with follow-up optical or NIR data, have been carried out by Zwaan et al. (1997), 
Spitzak and Schneider (1998), Rosenberg and Schneider (2000), Minchin et al. (2003),  Davies et al. (2004), 
Giovanelli et al. (2005). Some of the results are discussed in Rosenberg et al. (2005), Rosenberg and 
Schneider (2003) and Minchin et al. (2004). 
The main results can be summarised briefly as follows:
\begin{enumerate}
\item all but one galaxy (Minchin et al. 2005, 2007) seem to have optical counterparts;
\item although spirals predominate, the counterparts cover a wide variety of morphological types and luminosities;
\item there are significant, but not overwhelming, numbers of low surface brightness galaxies;
\item to within the measurement errors all \hi galaxies have the same \hi column densities 
($\sim 10^{20.65 \pm 0.38} cm^{-2}$) and this is not a selection effect.
\end{enumerate}
The various surveys differ from one another partly in radio sensitivity, but 
mainly in the quality of their optical or NIR follow-up data. Our Equatorial Survey is distinct in two main 
respects: a) we have been fastidious in rejecting all sources where there might be more than one galaxy in 
or near the beam; b) we have available high quality multi-band SDSS-DR2 data for every source, data which is both 
sensitive and uniform, across a wide dynamic range. Rosenberg at al.
(2005) have attempted a similar analysis using, 
instead of optical, 2MASS J-band data.

Our most interesting results can be summarised as follows: 
a) All galaxies have the same mass-to-light ratio in H-band solar units.  
In other words Gavazzi's Law holds over a dynamic range of 8 mag in absolute luminosity 
even in \hie-selected galaxies free of optical selection effects. 
b) There is a clear correlation between SB and Luminosity. Combined
with (a) it reveals a correlation between dynamical mass and optical
radius cubed, again over a large dynamic range, but now with more
scatter. In other words there is roughly constant global-density for
galaxies (\mbox{$\sim$ 1 $\times$ 10$^{-24}~gm~cm^{-3}$}). The Virial Theorem
then implies the same angular velocity \mbox{$\sim$ 4 $\times$ 10$^{-16}
~radians~s^{-1}$}, implying they could have rotated no more than 40
times in a Hubble epoch.  
(c) All the galaxies have the same global surface density in \hie,
that is to say the same \hie-mass divided by optical-radius squared (R$^{2}$); 
(d) All the galaxies have exponential profiles throughout most of
their extents so that $R_{90}$ is correlated very tightly with $R_{50}$; 
(e) There is a clear colour-luminosity correlation in the sense that more luminous galaxies are redder, which is 
well known in optically selected samples. 

The remainder of this paper is arranged by section as follows:
     
(2) \textit{The Radio Observations} describe the \hi survey, the
detection process and the contents of the resulting catalogue.

(3) \textit{The Optical Observations} looks into the sensitivity of the SDSS for picking up large angular-size 
low surface brightness galaxies, and galaxies with low $\left(\frac{M_{H\!I}}{L_{B}}\right)$ ratios. It then discusses 
the general problem of identifying reliable optical counterparts to \hi sources in blind \hi surveys. 
 
(4) \textit{The Diversity of Sources} displays complete data 
sets for a handful of detected objects, 
ranging from giant spirals to inchoate low surface brightness
galaxies, which represent important \hie-selected class types. 
A montage of the optical counterparts illustrates their very 
wide variety. Finally the complete data set is laid out in the form of tables.

(5) \textit{The Correlations in Galaxy Properties} reveals the 5
separate correlations among their properties.

(6) \textit{The Discussion} argues that the correlations, despite the small size of the sample, are significant, 
not unduly affected by distance uncertainties, and compatible with the Tully-Fisher relation: Line-width $\sim$ 
L$^{\alpha}$, provided $\alpha$ $\simeq$ 1/3. If no more than 6 physical invariants control galaxies, as we argue, 
then 5 correlations suggest a degree of organisation among gas-rich galaxies which is surprising, and which must 
strongly constrain theories as to their formation and evolution.

\section{THE RADIO OBSERVATIONS}

The observations were carried out with the Parkes 64-meter radio telescope using the Multibeam System 
(Staveley-Smith et al. 1996) in which 13 adjacent beams employing 26 receivers track the sky at a rate 
of 1 degree per minute, returning several times to the same piece of sky weeks apart to minimise 
interference. The main characteristics of the survey are listed in Table~\ref{hipass}. 

\begin{figure}
 \begin{center}
  \subfigure[\label{1a}Distribution of R.A.'s]{\includegraphics[width=8.4cm]{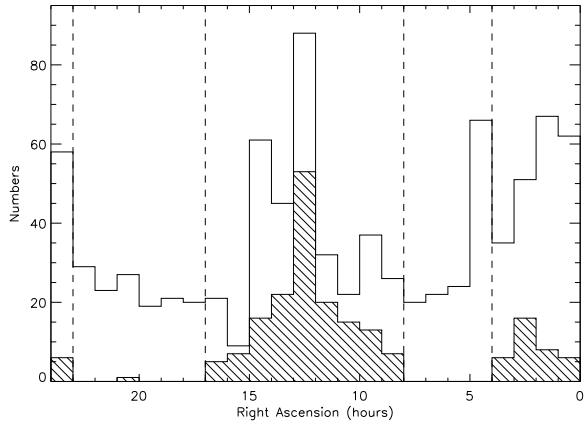}}
  \subfigure[\label{1b}Distribution of Declinations]{\includegraphics[width=7.7cm]{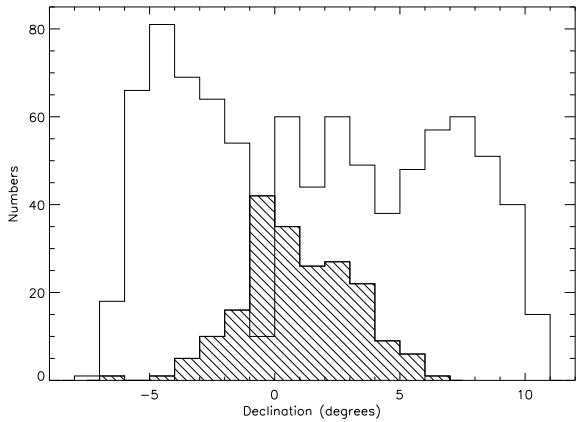}}
  \caption{\label{spadist} Spatial distribution of the ES sample. 
(a) shows the distribution in right ascension. The dashed lines 
indicate the range where there is no optical SDSS-DR2 data available. 
(b) shows the distribution of ES sources in declination. The unfilled 
histogram indicates those \hi sources without optical SDSS-DR2 data 
and the line-filled histogram the sources with optical SDSS-DR2 data.}
 \end{center}
\end{figure}

\begin{figure}
 \begin{center}
  \subfigure[\label{2a}Distribution of peak fluxes 
  ($S_{peak}$)]{\includegraphics[width=7.9cm]{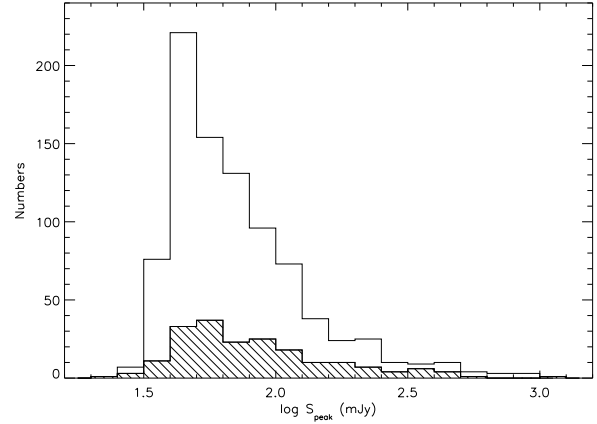}}
  \subfigure[\label{2b}Distribution of integrated fluxes 
  ($S_{int}$)]{\includegraphics[width=7.9cm]{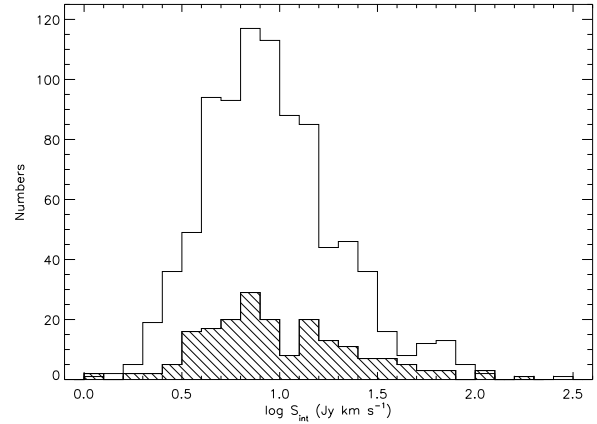}}
  \caption{\label{peak} Flux distributions for the ES sample. 
The unfilled histogram indicates those \hi sources without 
optical data and the line-filled histogram the sources 
with optical SDSS-DR2 data.}
 \end{center}
\end{figure}

The HIPASS catalogue, that is to say the pipeline catalogue of sources based on this raw data,  has been 
released over the years using techniques, and with results described in Barnes et al. (2001), Meyer et al. (2004),  
Zwaan et al. (2004) and Wong et al. (2005). Our Equatorial Survey, based on the same raw data, but covering only 
5700 deg$^{2}$ around the Equator (Garcia-Appadoo, PhD Thesis, 2005) was analysed much 
earlier (1999) with the intention of providing an early source list in time for SDSS-DR2 release.  Although it 
used almost identical search methods to HIPASS, and indeed copies most
of them directly, it appears marginally more sensitive than the HIPASS catalogue of the
same region (Wong et al. 2005), perhaps because we  
had more time to follow up marginal detections with the more sensitive narrow-band system at Parkes 
(Zwaan et al. 2004) thus winning fainter signal from noise. 
The Equatorial Survey (ES hereafter) covers 14 per cent of the entire sky in a band right round the celestial 
equator from declination -6\arcdeg to +10\arcdeg. Of this area about 1700 deg$^{2}$ is covered by the 2nd Data Release 
of the SDSS (Abazajain et al. 2004) which we will use as our source of optical information. 
Fig. \ref{spadist} shows the distribution of sources around the sky, 
both those with SDSS-DR2 optical data and those without. The gaps 
caused by the Galactic Plane, as well as the strong peak at an R.A. 
of $\sim$ 12 hours due to the southern extension of the Virgo cluster 
are easily visible, note that the source at 20h R.A. lies in an additional 
small area of SDSS-DR2 coverage that, due to its small size, is not shown 
in Fig. \ref{spadist}. Fig. \ref{peak} shows the 
distributions of ES sources in both peak and integrated flux. 
As can be seen the distributions for both the ES sample with (hashed) and 
the sample without (clear) optical data are very similar. 
This indicates that the optical sub-sample we will be analysing here is 
representative of the sample as a whole and hence can be used to characterise 
the properties of \hi selected galaxies in general.

\begin{small}
\begin{table}
  \centering
  \begin{tabular}{lc}
    \hline
    Parameter & \hipass value \\
    \hline
    Sky Coverage & $\delta~<$ +25\arcdeg\\
    Integration time per beam & 450 s\\
    Average FWHM & 14.3 arcmin\\
    Gridded FWHM & 15.5 arcmin\\
    Pixel size & 4 arcmin\\
    Velocity range & $-$1280 to 12700 \kms\\
    Channel Separation & 13.2 \kms\\
    Rms noise & 13 mJy beam$^{-1}$\\
    3$\sigma$ \hi Mass Limit$^{a}$ & 10$^{6}$ D$^{2}_{Mpc}$ \msun\\
    N$_{H\!I}$ limit$^{a}$  & 7.8 $\times$ 10$^{18}$ cm$^{-2}$\\
    \hline
    $^{a}$ For $\Delta$V = 100 \kms \\
  \end{tabular}
\caption{\label{hipass} Summary of \hipass survey parameters}
\end{table}
\end{small}

The noise in the Parkes Multibeam data is complex, consisting of a mix of receiver noise, solar side-lobe emission, 
confusion and ripple in the 21-cm spectra with a characteristic frequency of $\sim$ 1200 \kms due to standing waves 
with the wavelength of 52 meters set up between the dish surface and the prime focus cabin by continuum sources 
in the beam.  With such complex noise the selection effects which enter into the discrimination of weak sources 
cannot be anticipated, and must be recognised retrospectively. 
Multibeam data is presented to the observer in cubes with axes in the R.A., declination and radial-velocity 
directions. The cubes are 8 degs on a side separated into 4 arcmin pixels and divided in the third dimension 
into 1024 velocity channels each 13.1 \kms wide. Given that the original radio beam was 7 arcmin 
half-maximum in diameter the pixel signals are partially correlated, as are the velocity channels which have 
been Hanning-smoothed to 18 \kmse. These cubes (102 of them in our case) were searched individually 
using a mixture of numerical algorithms and the eye-brain system.  Algorithms are helpful, particularly 
in reducing labour; but none has so far proved anywhere as reliable or sensitive as the eye-brain 
(e.g. Kilborn 2002). Each cube can be represented in 3 combinations 
($\alpha$, $\delta$), ($\alpha$, V) and ($\delta$, V) where V is the radial velocity, and all 3 are searched before deciding 
a source is probably present. Such probable sources can be followed up by using the 
multibeam system in a more sensitive narrow-band mode. A combination of much higher velocity resolution 
(1.3 \kms per channel) and the ability to integrate indefinitely on a possible source, rather than for a total 
of 450 seconds in the scans, is a powerful filter against spurious misidentifications (Zwaan et al. 2004). 
Note that the existence, or otherwise, of an optical counterpart is in no sense used in selecting or rejecting a 
source, making the survey truly blind.

The ES \hi sample was selected using methods almost identical to
HIPASS selection (Meyer et al. 2004, Zwaan et al. 2004). The noise is
complex so the only way to establish the likely reality of sources is
to follow up a sufficient proportion with the more sensitive
`Narrowband System'. The hundreds of putative HIPASS sources followed up
gives confidence that significantly less than 5\% of the remaining ES
sources could be spurious, far too small fraction to affect the
correlations we are searching for [Section 5].

\begin{figure}
 \begin{center}
   \includegraphics[width=8cm]{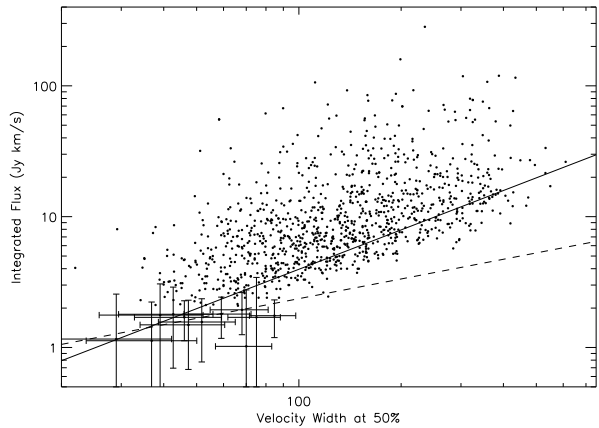}
   \caption{\label{selint} Selection limits in Velocity Width-integrated
   flux space. The theoretical 3$\sigma$ limit for selection based on S$_{int}$ (constant signal-to-noise) is
   shown by the dashed line and the 3$\sigma$ limit for S$_{peak}$ selection ($\frac{S_{int}}{\Delta V_{50}}$) is
   shown by the solid line. Most error bars have been omitted for clarity.}
  \end{center}
\end{figure}

\begin{figure}
 \begin{center}
   \includegraphics[width=8cm]{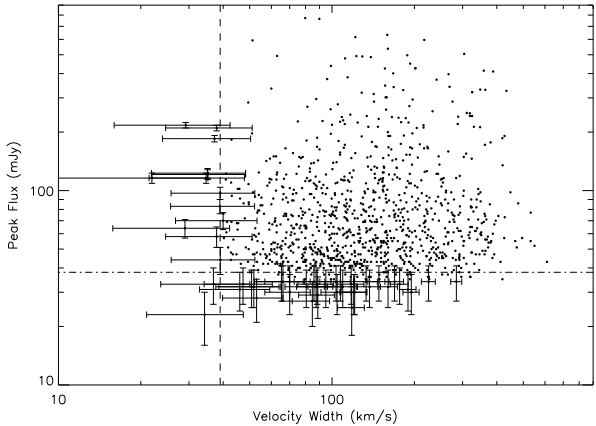}
   \caption{\label{selpeak} Selection limits in velocity width-peak flux space. The 3$\sigma$ (39 mJy) shown here (dot-dashed line) 
   can be seen to be a good match to the selection limit of the data. The dashed vertical line shows $\Delta V_{50}$ = 3 
   channels (39.6 \kmse). Most error bars have been omitted for clarity.}
  \end{center}
\end{figure}

Fig. \ref{selint} shows the integrated flux as a
function of the velocity width of the sources. It appears that 
detection by integrated flux is dependent on line width, with 
broad line sources being harder to find in the noise. In peak flux however 
it is much easier to set a clean selection criterion independent 
of line-width (see Fig. \ref{selpeak}), as Kilborn (2002) and 
other searchers in the multibeam data have found. 

\begin{figure}
 \begin{center} 
   \subfigure[\label{5a}Numbers of the ES sample by peak flux ($S_{peak}$)]{\includegraphics[width=8cm]{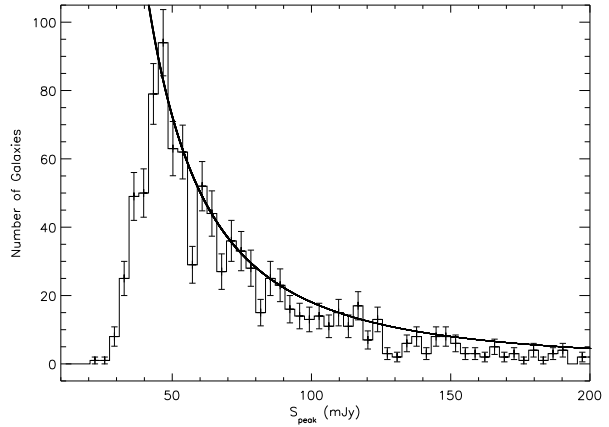}}
   \subfigure[\label{5b}Numbers of the ES sample by integrated flux ($S_{int}$)]{\includegraphics[width=8cm]{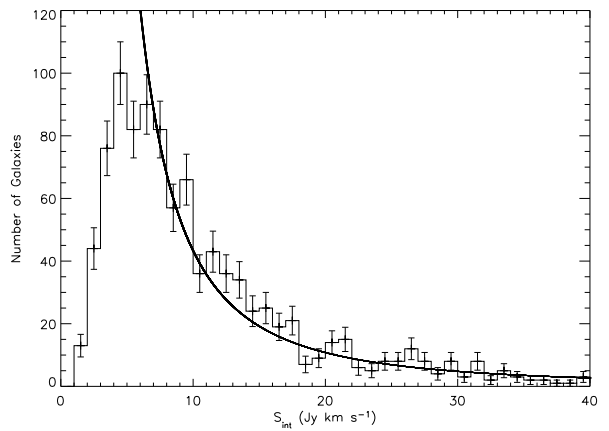}}
   \caption{\label{complet} (a) Source count against peak flux. The histogram shows the numbers found in
    each bin of peak flux. (b) the completeness of the ES sample in integrated flux. The histogram shows the numbers 
    found in each bin of peak flux. The curve in both panels represents \emph{$N(S)\sim S^{-5/2}$} as expected for a flux-limited survey.}
    \end{center}
\end{figure}

Fig. 5 shows flux limited curves $N(S) \sim S^{-5/2}$ filled to peak
fluxes (a) and integrated fluxes (b). Fig. \ref{5a} is suggestive of
a peak-flux completeness limit $\sim$ 50 mJy.  

\begin{figure}
 \begin{center}
  \subfigure[\label{6a}distribution of recessional velocities]{\includegraphics[width=8cm]{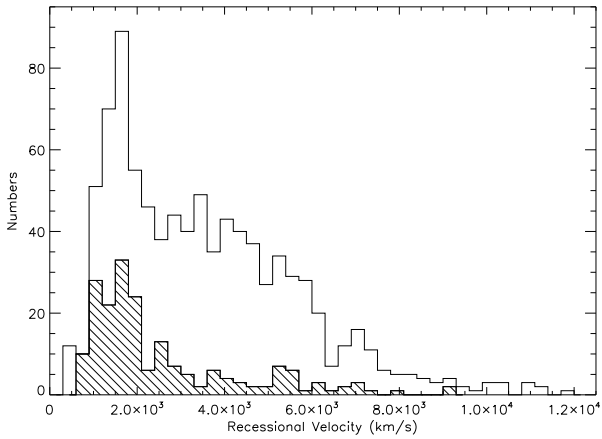}}
  \subfigure[\label{6b}distribution of velocity widths]{\includegraphics[width=8cm]{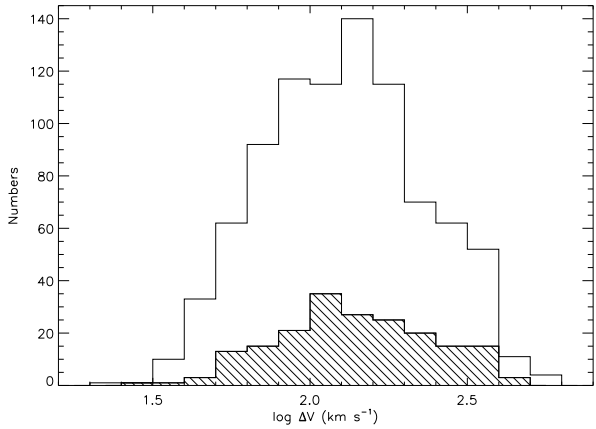}}
  \caption{\label{vels} (a) shows the recessional velocity distribution of the ES sample. (b) shows the 
  distribution of the 50\% velocity widths ($\Delta V_{50}$) for the ES sample. The unfilled histogram indicates 
  those \hi sources without optical data and the line-filled histogram the sources with optical SDSS-DR2 data.}
 \end{center}
\end{figure}

The raw velocities are corrected for solar motion relative to the
Local Group. The resulting distribution of radial velocities is shown 
in Fig. \ref{vels}. Approximate distances are derived in a similar way
as in Koribalski et al. (2004), using \mbox{$D = v_{LG}/H_{0}$}, where
\mbox{$v_{LG} = v_{sys} + 300~sin~l~cos~b$}, H$_{0}$ = 75 km s$^{-1}$ Mpc$^{-1}$ 
is assumed throughout (and where necessary a cosmology with 
$\Omega_{M}$ = 0.3 and $\Omega_{\Lambda}$ = 0.7).

\begin{figure}
 \begin{center}
  \includegraphics[width=8cm]{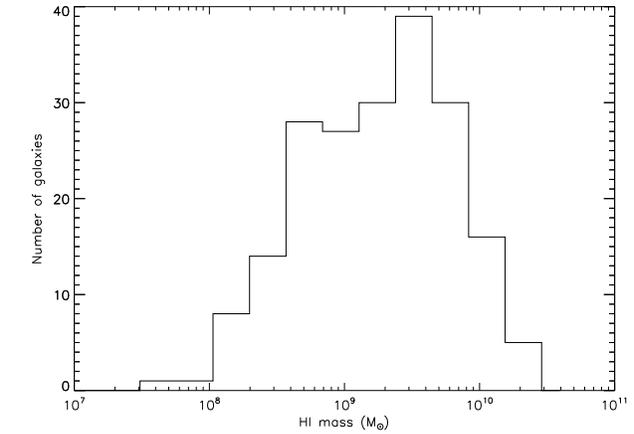}
  \caption[\hi mass distribution for the ES sample]
  {\label{HImass} Distribution of \hi masses for the ES sample}
   \end{center}
\end{figure}

The \hi masses were derived from the integrated fluxes using fitting 
processes identical to HICAT and:
\begin{equation}
M_{HI} (M_\odot) = 2.36 \times 10^{5} \times D(Mpc)^{2} \times \int S(H\!I)~dV 
\end{equation}
where $D$ is the distance in Mpc and the integral $\int S(H\!I)dV$ is 
the integrated flux in Jy \kmse, equation (3) assumes that all the
sources are optically thin and that the upper levels of the 21-cm 
transition are fully excited. Fig. \ref{HImass} shows the distribution 
of \hi masses for the ES sample.
\begin{table*}
\centering
\begin{minipage}{146mm}
\caption{\hi properties of the ES sample\label{HIdatatable}}
\begin{tabular}{lrrccrcrr}
\hline \\[-2.5ex]
\multicolumn{1}{c}{ES Name} & 
\multicolumn{1}{c}{RA}  &
\multicolumn{1}{c}{Decl.}  & 
\multicolumn{1}{c}{$S_{peak}$} &
\multicolumn{1}{c}{$S_{int}$} & 
\multicolumn{1}{c}{v$_{sys}$} &
\multicolumn{1}{c}{W$_{20}$} & 
\multicolumn{1}{c}{D} &
\multicolumn{1}{c}{log M$_{\hi}$}      \\
& \multicolumn{1}{c}{(\small{J2000)}}
&\multicolumn{1}{c}{(\small{J2000)}}  
& \multicolumn{1}{c}{(Jy)} 
& \multicolumn{1}{c}{(Jy\,\kmse)} 
& \multicolumn{1}{c}{(\kmse)}
& \multicolumn{1}{c}{(\kmse)} 
& \multicolumn{1}{c}{(Mpc)}    
& \multicolumn{1}{c}{(M$_{\odot}$)} \\
\multicolumn{1}{c}{(1)} & 
\multicolumn{1}{c}{(2)} & 
\multicolumn{1}{c}{(3)} & 
\multicolumn{1}{c}{(4)} &
\multicolumn{1}{c}{(5)} &
\multicolumn{1}{c}{(6)} & 
\multicolumn{1}{c}{(7)} & 
\multicolumn{1}{c}{(8)} & 
\multicolumn{1}{c}{(9)} \\
\\[-2.0ex]
\hline
\\[-2.0ex]
HIPEQ0014-00 &   00 14 36 &   -00 44 42 &   0.075 &  17.18 & 3914 & 290 &
 50.9 & 10.02 \\
HIPEQ0027-01a &  00 27 47 &  -01 09 39 & 0.039 &   6.60 & 3848 & 223 &   54.2 &  9.66 \\
HIPEQ0033-01 &   00 33 22 &  -01 07 01 & 0.131 &  17.24 & 1972 & 146 &   30.1 &  9.57 \\
HIPEQ0043-00 &  00 43 31 &   -00 06 49 & 0.064 &  13.81 & 4124 & 287 &   60.6 & 10.08 \\
HIPEQ0051-00 &  00 51 57 &  -00 28 25 & 0.117 &  14.84 & 1616 & 173 &   26.0 &  9.37 \\
HIPEQ0058+00 & 00 58 50 &   00 37 46 & 0.048 &   4.97 & 5338 & 156 &   75.4 &  9.82 \\
\hline
\end{tabular}
NOTE.-- An extract of the table is shown here for guidance. It is
presented in its entirety in the electronic edition of the Journal.
\end{minipage}
\end{table*}

\begin{figure}
 \begin{center}
  \includegraphics[width=7cm]{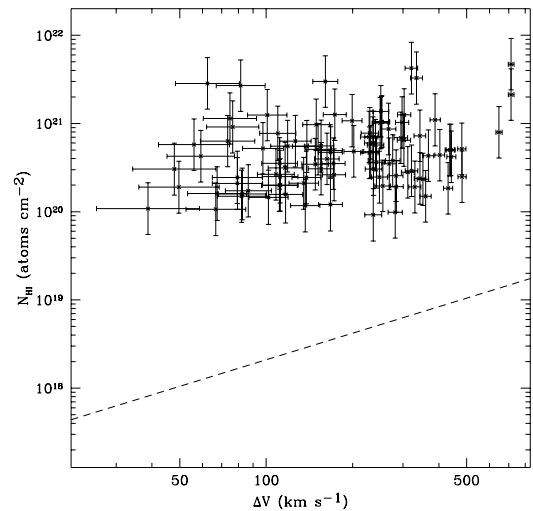}
  \caption{\label{NHI} The distribution of the N$_{H\!I}$ column densities among the $\sim$100 HIDEEP galaxies
observed by Minchin et al. (2003). The dashed line shows the sensitivity limit. There appear to be no low column density objects}
   \end{center}
\end{figure}

Although in this paper we are not going to try and compensate 
for \hi selection effects we need to be aware of how they may 
have shaped our sample. The first worry concerns 
column-density-sensitivity. Will we have sufficient sensitivity 
to pick up LSBGs with correspondingly low $N_{H\!I}$ 
(see equation (2))? It can be shown (Minchin et al. 2003) that the 
column-density-sensitivity, N$_{H\!I}$ (cm$^{-2}$) , of any survey 
can be worked out retrospectively from the fluxes F (Jy \kmse) and 
sizes $\delta \theta$ (\hi diameter in arcmin) of the faintest 
sources within it  according to:
\begin{equation}
\left[N_{H\!I}\right]_{min} = 4.5\times 10^{20}
\left(\frac{F{H\!I}}{\Delta
    V~\delta\theta^2}\right)_{min}^{gal}\times~\Delta V~\frac{atoms~km}{cm^{2}~sec}
\end{equation}
where $\Delta V$ is their velocity width in \kmse. 
This is the \hi equivalent of equation (1) in the optical. 
For the ES it corresponds to $\sim$ 10$^{19}$ cm$^{-2}$, or a low SB
limit, according to equation (2), where $\overline{\mu}_{21}$ is the
mean SB inside the 21-cm area of the galaxy:
\begin{equation}
\overline{\mu}_{21} (B) = 27 - 2.5 \log\left[\frac{N_{H\!I}}{10^{20.1}}~/~\left(\frac{M(H\!I)}{L_{B}}\right)\right]~\frac{mag}{arcsec^{2}}
\end{equation}
For a typical ($M_{H\!I}/L_{B}$) $\sim$ 1 this corresponds to a
central SB $\mu_{0}$ (B) $\sim$ 25.7 or an effective SB 
$\mu_{ef\!f}$ (B) $\sim$ 27.5 which is very dim, dimmer than 
any optical catalogue covering a significant area (see Table
\ref{definitions} for definition of different SB's). 
Indeed it is so dim that one might question whether the SDSS 
could reach it. But here enters a surprising result recently 
discovered by Minchin using the Parkes multibeam to carry out a much deeper \hi survey 
(HIDEEP; 9000 s beam$^{-1}$) of a small area, and with full optical follow up (Minchin et al. 2003). 
Despite its great sensitivity to low column densities ($N_{H\!I}\geq 2 \times 10^{18}$ atoms cm$^{-2}$) 
they found that there are  no low column density galaxies. Indeed they found that all \hi selected galaxies have, 
to within the errors, the same column density [$\sim$ 10$^{20.5}$ cm$^{-2}$ if spread over 5 effective radii, 
as in optically selected samples (Salpeter \& Hoffman 1996)], see Fig. \ref{NHI}.
We find the identical uniformity of $N_{H\!I}$ in the ES, see Fig. \ref{11a}.

Whatever the reason for Minchin's strange law (Minchin et al. 2003), and it needs explaining, the dimmest LSBGs 
we can expect to encounter in the ES will have, according to equation (5), and the fact 
that \mbox{$\overline{\mu}_{ef\!f} = \overline{\mu}_{21} -2.1$} mag for an exponential disc:
\begin{equation}
\mu_{ef\!f} (B) (mag) = 23.9 + 2.5 \log\left(\frac{M(H\!I)}{L_{B}}\right)
\end{equation}
which should be comfortably accessible with SDSS for galaxies with $(M_{H\!I}/L_{B})_{\odot}$ $<$ 5 (see Section 3).

The \hi properties obtained for the ES sample are listed in Table
\ref{HIdatatable}, which is provided in full in the electronic edition
of this journal. The columns are as follows:

\textit{Column (1).}-- ES source name.\\
\indent \textit{Column (2) and (3).}-- Fitted \hi positions in right ascension
and declination (J2000).\\
\indent \textit{Column (4).}-- \hi peak flux density, S$_{peak}$.\\
\indent \textit{Column (5).}-- Integrated \hi flux density, S$_{int}$.\\
\indent \textit{Column (6).}-- \hi systematic velocity, v$_{sys}$, measured at
the 20\% level of peak flux density.\\
\indent \textit{Column (7).}-- Velocity line width, W$_{20}$, measured
at 20\% level of the peak flux.\\
\indent \textit{Column (8).}-- Distance, D, in Mpc (see text for description). \\  
\indent \textit{Column (9).}-- The logarithm of the \hi mass,
M$_{\sc{H\!I}}$, calculated using equation (3) in units of solar masses. 

\section{OPTICAL IDENTIFICATION AND OBSERVATIONS}

\begin{table*}
\centering

\caption{Definitions of surface brightness for galaxies with pure exponential profiles\label{definitions}}
\begin{tabular}{ll}
\hline
\textbf{Surface Brightness measurement} & \textbf{Definition}\\
\hline
$\mu_{0}$ & Central Surface Brightness (SB)\\
$\mu_{50}$~(= $\mu_{0}$ + 1.82$^{m}$) & The SB \textit{at} the half-light radius (sometimes called `effective SB')\\
$\overline{\mu}_{50}$~(= $\mu_{0}$ + 1.12$^{m}$) & The mean SB \textit{within} the half-light radius (also called `effective SB')\\
$\mu_{90}$~(= $\mu_{0}$ + 3.9$^{m}$) & The SB \textit{at} the 90-per-cent-light radius\\
$\overline{\mu}_{90}$~(= $\mu_{0}$ + 2.3$^{m}$) & The mean SB \textit{within} the 90-per-cent-light radius\\ 
$\overline{\mu}_{21}$~(= $\mu_{0}$ + 3.84$^{m}$) & The mean SB \textit{within} the outermost 21-cm contour (see text)\\ 
\hline
\end{tabular}\\
NOTE.-- $\mu_{0}$ in magnitudes corresponds to the central SB I$_{0}$ where $I(r) = I_{0} e^{-r/\alpha}$ and $\alpha$ is the 
scale length.\\ Their total luminosity $L_{T} = 2\pi I_{0} \alpha^{2}$; R$_{50}$ = 1.68 $\alpha$, R$_{90}$ = 2.32 R$_{50}$ 
and (empirically) R$_{21}$ $\approx$ 5 R$_{50}$ = 8.4 $\alpha$.
\end{table*}

We determined that the best source of optical data to follow up the radio survey would be the SDSS. 
With its 5 photometric bands ($u, g, r, i, z$), it offers a unique and homogeneous data-base. The SDSS 
was however initially aimed at working on small faint objects and the pipeline software was not designed 
to cope with the arcminute sized galaxies which turn up in the ES so that up to now most SDSS galaxy 
work has had to be confined to objects beyond z $=$ 0.02 (e.g. Blanton et al. 2003). 
Identifying the various large galaxy problems, finding and validating solutions for them, and rewriting parts of the software 
has held up the ES for at least 2 years. This is no place to discuss those corrections as an account appears 
in West et al. (2008). We summarise them only briefly to emphasise how necessary they are when 
using SDSS to study galaxies of arcminute size.

\begin{figure}
\centering
\includegraphics[width=7cm]{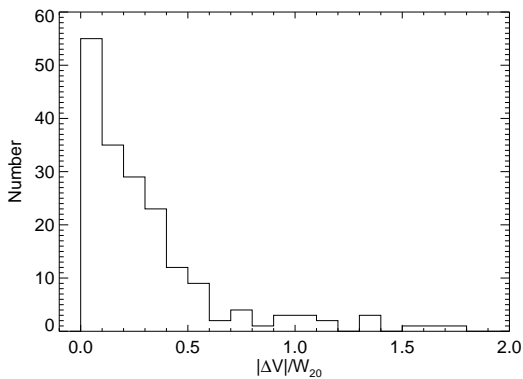}
\caption{\label{DVW} Distribution of velocity differences $\Delta V/W_{20}$ between radio sources and their 
optical counterparts in the ES for the original 310 sources.}
\end{figure}

\subsection{Limiting surface brightness}
One legitimate concern about the SDSS, which is a relatively shallow survey ($\sim$ 55 s exposure per point), 
is whether it will go deep enough to detect the kind of LSBGs one might hope to find in the ES. 
Tests show that the main source of noise in looking for dim extended objects in the SDSS is photon 
noise in the sky subtraction (West et al. 2008, Strateva et al. 2001). In that case it can be shown that:

\begin{equation}
f(a) \geq \left(\frac{S}{N}\right) \frac{1}{\theta^{''}} \frac{1}{\sqrt{N(a)}}
\end{equation}

where f(a) is the fraction of the sky level one can go down to and find objects of angular size $\theta^{''}$ (diameter in arcsec) with a given signal-to-noise f(a) is that fraction for filter a, while N(a) is the total number of photons accumulated 
from the sky arcsec$^{-2}$ through filter a. For the $g$-band SDSS N(g) $\sim$ 120, and if one demands a S/N of at 
least 10 for detection:

\begin{equation}
\overline{\mu}_{21}(g) \leq 22.1 + 2.5 ~ log_{10} ~ \theta^{''}
\end{equation}

For mapped galaxies the $\theta^{''}_{21}$ is 
typically 10 $\theta^{''}_{50}$ where $\theta^{''}_{50}$ is the half light radius (e.g. Salpeter \& Hoffman 1996). 
As $\overline{\mu}_{21}$ = $\mu_{0}$ + 3.84, detection with SDSS in the $g$ band requires:

\begin{equation}
\mu_{0}(g)~\leq~20.8~+~2.5~log_{10}~(\theta^{''}_{50})
\end{equation}
In contrast LSBGs are generally defined to have $\mu_{0}(B) <$ 23.0 (see below). In other words, as you would 
expect of such a short exposure survey, the SDSS will be capable of picking up LSBGs only if they have large 
angular size ($\theta_{50} >$ 10 arcsec).  
In the case of the ES all the sources are nearby ($\leq$ 10,000 \kms away) and although some of our 
claimed identifications lie close to the SB detection limit (18),  they appear visible because they 
have patches of light that are brighter than average,  such as HII regions. Of course we will not
detect galaxies with relatively small fractions of \hie,  dEs and dSphs for instance, unless the galaxies 
are very massive.  We can  quantify 
this limitation by combining (9) with (5) in which case,  for typical values of 
\mbox{log N$_{H\!I}$ $\sim$ 20.5 cm$^{-2}$} (Minchin et al. 2003) and
sky brightness, $\mu_{sky}$(g) = 22.5 g mag arcsec$^{-2}$, the ES will 
only detect galaxies for which:

\begin{equation}
\theta^{''}(g) / (M_{H\!I}/ L_{B})_{\odot} \geq 15^{''}
\end{equation}

\ni where $\theta^{''}$ is the optical diameter (corresponding to 2 $\times$ $\theta_{90}$) in the $g$ band.  
Gas poor galaxies will only be found if they are close by (i.e. $\theta^{''}$ large).
In other words the ES may miss a significant fraction of light in the
Universe coming from gas-poor galaxies. It is no coincidence that all 
the galaxies we do detect appear to be late-type.
       
\begin{figure}
\centering
\includegraphics[width=7cm]{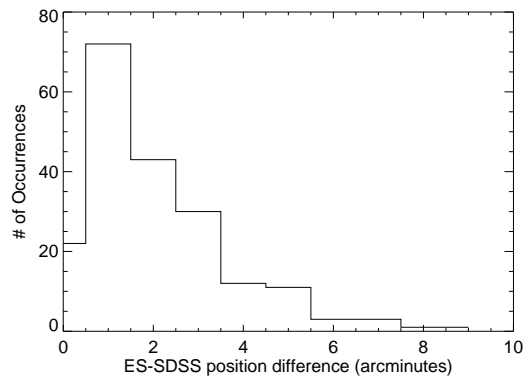}
\caption{\label{posdist} Distribution in positional differences between the radio and optical sources for the final list of 195 sources.}
\end{figure}
  
Many definitions of galaxy SB, and what constitutes low SB,  appear in the literature. For instance 
there are two quite different definitions of effective SB that are current: $\mu_{50}$, which is the 
actual SB at the half-light radius, and $\overline{\mu}_{50}$ which is the mean value of the SB \textit{within} the 
half-light radius. To avoid even further confusion we list all these definitions together in Table \ref{definitions}
and relate them all to the central SB $\mu_{0}$. The relationships are only valid for pure exponential profiles.  
Where we have converted from SDSS colors to B-V system we have used Cross et al. (2004).

What constitutes a low surface brightness galaxy must be a matter of convention. However most optically selected surveys 
show a distribution  in SB, however defined, which is approximately Gaussian with a FWHM $\sim$ 3.0 magnitudes. 
Thus galaxies with a SB 1.5 or more magnitudes dimmer than the peak value could reasonably be defined as LSBG's
and that seems to be a common,  though not universal, convention (e.g. Impey \& Bothun 1997). For photographic, 
and for shallow CCD surveys, the peak in the disc SB distribution appears to lie close to the Freeman value 
of \mbox{$\mu_{0}$(B) = 21.65}. Thus any disc dimmer than \mbox{$\mu_{0}$(B) = 23.0 $\mu$(B)} could reasonably be regarded as LSB. 
The median SB for the ES, allowing for average bulge-to-disk ratio of $\sim$ 0.1, is $\mu_{0}(B)\sim$ 22.5 mag arcsec$^{2}$, 
or about 0.8 mag arcsec$^{2}$ dimmer than the Freeman value for optical surveys, and we find objects as dim as 
$\mu_{0}(B)\sim$ 24 mag arcsec$^{2}$.

\subsection{Identification of optical counterparts}
The unambiguous identification of optical counterparts to 21-cm sources found in a blind survey like 
ours is by no means trivial. Because of the strong clustering of galaxies,  both in angular and in 
redshift space, it is all too easy to find a plausible optical counterpart for virtually every 21-cm 
source. It is surprising, for instance, that in HOPCAT,  the published optical catalogue for the 
4315 HIPASS sources in the Southern sky (Doyle et al. 2005) there is not one source,  not one 
intergalactic gas cloud or dark galaxy without a plausible 
optical counterpart. We shall now estimate the rate of false identification, i.e. the probability of 
finding a random optical galaxy within a given distance, in both angular and redshift space, of any 
given \hi source.  We shall assume, as the observations clearly suggest (Staveley-Smith et al. 1996) 
that optical galaxies and \hi sources are clustered together. 
For an ES or HIPASS source the acceptable volume V$_{acc}$ in which an optical counterpart could lie 
is a long thin cylinder, centred on the source,  with its long axis,  set by the radial-velocity 
uncertainties, along the line of sight.  For an ES source at a typical radial velocity of 2000 \kmse, 
the angular uncertainties in position (= R$_{0}$) (up to 5 arcmin) correspond to $\sim$ 50 kpc, while the 
velocity uncertainties $\Delta V$, taking account of both radio and optical uncertainties, amount to 
H$_{0}$ $\Delta V$ ($\sim$30 \kmse) ($\equiv$ h) or half a Mpc. Given the correlation function:

\begin{equation}
p(r) dV = n_{0} dV [1 + \xi(r)]
\end{equation}

\ni where \mbox{$\xi(r) = \left(\frac{r}{r_{0}}\right)^{-1.8}$} and $n_{0}$ is the average number of plausible galaxies Mpc$^{-3}$, 
it is possible to integrate the probability of finding a random galaxy within the volume V$_{acc}$ of the acceptable cylinder. 
To a very good approximation the number within a projected distance R$_{0}$ (in Mpc) of the source is given by

\begin{equation}
N(< R_{0}) \approx 1.8~n_{0}~r_{0}^{1.8}~R_{0}^{1.2} ~ (r_{0} \approx 8 Mpc)
\end{equation}

Notice that the number is only weakly dependent on R$_{0}$ (because of the  strong correlation) and dependent 
on the radial-velocity uncertainty not at all. This last is counter-intuitive but arises from the long thin 
shape of the cylinder. The ends of the cylinder are so very far from the centre that finding highly 
correlated galaxies within the ends is very unlikely. Conversely, obtaining very accurate optical velocities 
for plausible galaxies in the field does not greatly enhance ones chances of making an unambiguous 
identification when the \hi velocity-uncertainties may still (for S/N reasons) be larger (and oddly enough,  
making blind \hi surveys with bigger dishes will not help because the characteristic sources will be 
proportionately farther away (Disney 2008). 

To turn equation (12) into numbers it is necessary to adopt an optical Luminosity Function for the putative galaxies.
If we adopt:

\begin{equation}
\varphi(M) =  \varphi_{*} \kappa^{\alpha+1} e^{-\kappa}
\end{equation}

\ni where \mbox{$\kappa \equiv 10^{0.4(M_{*}-M)}$}, $\alpha$ = 1.2, $\varphi_{*}$ = 2.1 $\times$ 10$^{-2}$ h$^{-3}$ 
and \mbox{M$_{*}$ = $-$20.04 $-$ 5 log h} (Blanton et al. 2003) then
Table \ref{random} where Col. (1) gives the angular size distance in Mpc, 
R$_{0} = D~\Delta \theta$, $D$ is the source distance in Mpc,
$\Delta \theta$ is the angular distance between the source and the
optical counterpart in radians and Col. (2) is the number of random 
galaxies (down to 3 mag below M$_{*}$) to be expected within that distance.
As an example, consider a source at a typical
radial-velocity-distance of 2000 \kms in ES ($\sim$ 27 Mpc) 
where 1 arcmin corresponds to $\sim$ 8 kpc. The typical positional 
uncertainty in HIPASS $\sim$ 1.3 arcmin (Meyer et al. 2004, Zwaan et al. 2004) 
corresponds to $\sim$ 0.01 Mpc. 
At that distance, according to Col. (2) of Table 4, there is a 10 per
cent chance of finding a random galaxy of roughly the right radial 
velocity, i.e. clustered with the \hi source, within the positional 
uncertainty. And the chance of a misidentification could be higher
still if one was prepared to consider, as plausible candidates, 
objects more than 3 mag below L$_{*}$ in the luminosity function 
(our assumption in table \ref{definitions}). Note that Doyle et al. 
(2005) identify optical candidates up to 5 arcmin away, leading to 
the probability of a plausible misidentification 
\textit{at that distance} of more than 1.

\begin{table}
\centering
\caption{Number of random galaxies within D (Mpc)~$\Delta \theta$ (rad)
  of radio position.\label{random}}
\begin{tabular}{cc}
\hline
\hline
\multicolumn{1}{c}{R$_{0}$} & 
\multicolumn{1}{c}{N($<R_{0}$)}\\
\multicolumn{1}{c}{(1)} &
\multicolumn{1}{c}{(2)} \\
\hline
2 & 78\\
1 & 35\\
0.5 & 13\\
0.3 & 8\\
0.2 & 5\\
0.1 & 2.1\\
0.07 & 1.3\\
0.05 & 0.8\\
0.03 & 0.4\\
0.01 & 0.1\\
\hline 
\end{tabular}
\end{table}

\begin{table*}
\centering
\begin{minipage}{17.5cm}
\caption{Petrosian Photometry of the ES sample\label{optdatatable}}
\begin{tabular}{@{}lrrrrrrrrr@{}}

\hline \\[-2.5ex]
\multicolumn{1}{c}{ES Name} & 
\multicolumn{1}{c}{RA}  & 
\multicolumn{1}{c}{Decl.}  & 
\multicolumn{1}{c}{$u$} & 
\multicolumn{1}{c}{$g$} & 
\multicolumn{1}{c}{$r$} & 
\multicolumn{1}{c}{$i$} & 
\multicolumn{1}{c}{$z$} & 
\multicolumn{1}{c}{PetroR$_{50}$} & 
\multicolumn{1}{c}{PetroR$_{90}$} \\
 & \multicolumn{1}{c}{(J2000)} & 
\multicolumn{1}{c}{(J2000)} &  
&  
& 
&
 & 
&\multicolumn{1}{c}{($^{\prime\prime}$)} &
 \multicolumn{1}{c}{($^{\prime\prime}$)}\\
\multicolumn{1}{c}{(1)} & 
\multicolumn{1}{c}{(2)} & 
\multicolumn{1}{c}{(3)} & 
\multicolumn{1}{c}{(4)} &
\multicolumn{1}{c}{(5)} &
\multicolumn{1}{c}{(6)} & 
\multicolumn{1}{c}{(7)} & 
\multicolumn{1}{c}{(8)} &
\multicolumn{1}{c}{(9)} &
\multicolumn{1}{c}{(10)} \\
\\[-2.0ex]
\hline
\\[-2.0ex]
   HIPEQ0014$-$00 &    00 14 36 &   -00 44 42 &      $13.77\pm{0.04}$ &      $13.13\pm{0.02}$ &      $12.91\pm{0.02}$ &      $12.82\pm{0.02}$ &      $12.77\pm{0.03}$ &   $21.4\pm{0.4}$ &   $59.0\pm{0.8}$\\ 
  HIPEQ0027$-$01a &    00 27 47 &  -01 09 39 &      $14.72\pm{0.04}$ &      $14.09\pm{0.02}$ &      $13.85\pm{0.02}$ &      $13.79\pm{0.02}$ &      $13.87\pm{0.04}$ &   $21.4\pm{0.4}$ &   $43.6\pm{0.8}$\\ 
   HIPEQ0033$-$01 &    00 33 22 &  -01 07 01 &      $15.65\pm{0.06}$ &      $14.86\pm{0.02}$ &      $14.58\pm{0.02}$ &      $14.44\pm{0.02}$ &      $14.40\pm{0.05}$ &   $20.2\pm{0.4}$ &   $45.9\pm{1.6}$\\ 
   HIPEQ0043$-$00 &   00 43 31 &   -00 06 49 &      $13.90\pm{0.03}$ &      $12.95\pm{0.02}$ &      $12.52\pm{0.02}$ &      $12.31\pm{0.02}$ &      $12.14\pm{0.03}$ &   $13.9\pm{0.4}$ &   $36.4\pm{0.4}$\\ 
   HIPEQ0051$-$00 &   00 51 57 &  -00 28 25 &      $15.42\pm{0.03}$ &      $14.60\pm{0.02}$ &      $14.23\pm{0.02}$ &      $14.06\pm{0.02}$ &      $13.99\pm{0.03}$ &    $6.3\pm{0.4}$ &   $17.4\pm{0.4}$\\ 
   HIPEQ0058+00 &   00 58 50 &   00 37 46 &      $14.90\pm{0.03}$ &      $13.91\pm{0.02}$ &      $13.43\pm{0.02}$ &      $13.19\pm{0.02}$ &      $13.07\pm{0.03}$ &   $10.7\pm{0.4}$ &   $23.0\pm{0.4}$\\ 
\hline
\end{tabular}
NOTE.-- An extract of the table is shown here for guidance. It is
presented in its entirety in the electronic edition of the Journal.
\end{minipage}
\end{table*}

To reduce these potentially serious misidentification problems in ES we threw out all (90 out of 310) sources which 
looked, on inspection of the SDSS-DR2 fields, to have more than 1 plausible optical counterpart within the radio beam 
(FWHM $\sim$ 14 arcmin) and we obtained accurate optical velocities for all the rest using either those provided in NED, 
the SDSS-DR2 fibre 
or, for 20 galaxies, the Dual Imaging Spectrograph on the 3.5 metre telescope at Apache Point Observatory. All candidates 
with optical velocities discrepant from the \hi value by more than half the 21-cm line-width $\Delta V_{20}$ were discarded. 
In addition, a number of other sources were removed either because they extended across 2 or more SDSS-DR2 fields, or 
because there was a saturated foreground star within 1 arcmin.

Altogether of the original 310 HIPASS sources with plausible SDSS-DR2 galaxies within the HIPASS beam and at the 
right redshift (as defined above) 90 were thrown out because of multiplicity i.e. for there being more than 1 (up to 5) 
good SDSS-DR2 candidates, 20 were too extended and 5 too near a bright star. Fig. \ref{DVW} shows the distribution 
of velocity differences $\Delta V/W_{20}$ for the \textit{original} 310 candidates and Fig. \ref{posdist} shows the distribution 
in positional differences between the radio and optical sources for the 195 which remain in the final ES list. The tail 
in this plot is due to the \hi centroid not being centered in the optical source.

Most of the identifications fall within 2 arcmin of the radio 
position, consistent with the errors in those positions measured in the general HIPASS catalogue using 
interferometry (Meyer et al. 2004). Nevertheless there remain a tail of optical candidates up to 7 arcmin 
(half the FWHM beam) away from the radio centroid. Because of clustering (see Table 3) we can 
not rule out the possibility that a handful of sources (probably less than 10) remain misidentified. 
This is too small a number to invalidate the main results. Nevertheless we should acknowledge biases 
in our sample against galaxies in tight groups (too many in the beam), galaxies that appear very large 
(overlapping SDSS-DR2 fields), and dark galaxies or intergalactic clouds for which we will too easily find 
plausible, optically bright alternatives. Most of these biases are difficult to avoid and must exist to 
an equal or greater extent in other \hi selected blind samples. In particular the number of sources 
($\sim$ 30 per cent) discarded because of clustering within the beam highlights the difficulty of measuring 
\hi mass functions which are not somehow adjusted for confusion.

The data we used all came from SDSS Data Release 2 (Abazajian et al. 2004) using the pipeline from DR3 
(Abazajian et al. 2005) but no SDSS catalogue data was used, for reasons outlined below. DR2 covers 3324 
deg$^{2}$, about half of which overlaps the ES. SDSS pipeline photometry of large galaxies is very 
inaccurate for a number of reasons. The pipeline shreds large galaxies into a number of pieces, circular 
apertures are not appropriate, the inclination is not properly taken account of and worst of all sky 
subtraction can subtract much of a large galaxy away from itself. Thus we had to find ways of measuring 
the significance of all these problems and devise alternative methods  of handling the data. These are 
discussed at length in West (West PhD thesis) and are being published (West et al. 2008) so that 
the community wanting to use SDSS to work on nearby galaxies can make use of them.

Table \ref{optdatatable} lists the SDSS-DR2 optical data for the ES
sample.
\indent \textit{Column (1).}-- ES source name.\\
\indent \textit{Columns (2) and (3).}-- Right ascension and declination
(J2000) of the optical position of each SDSS-DR2 source.\\
\indent \textit{Columns (4) to (8).}-- Petrosian magnitudes in each
SDSS band $u,g,r,i,z$, Unlike the native SDSS-DR2 Petrosian
magnitudes, our Petrosian apertures are elliptical. The shape of each 
aperture is defined from two-dimensional Sersic fits and we use the 
$g$-band to measure both the shape and size of each Petrosian
aperture. For some galaxies in the sample, a Sersic profile was not 
a good fit. In these cases, a circular aperture was used.\\
\indent \textit{Columns (9) and (10).}-- The PetroR50 and PetroR90
values reported are the $g$-band semi-major axes that encompass 50 
per cent and 90 per cent of the total Petrosian flux respectively. 
All of the SDSS data are described in West et al. (2008).

\section{THE DIVERSITY OF SOURCES}

\begin{figure}
\centering
  \subfigure[\label{11a}Distribution of column densities]{\includegraphics[width=7cm]{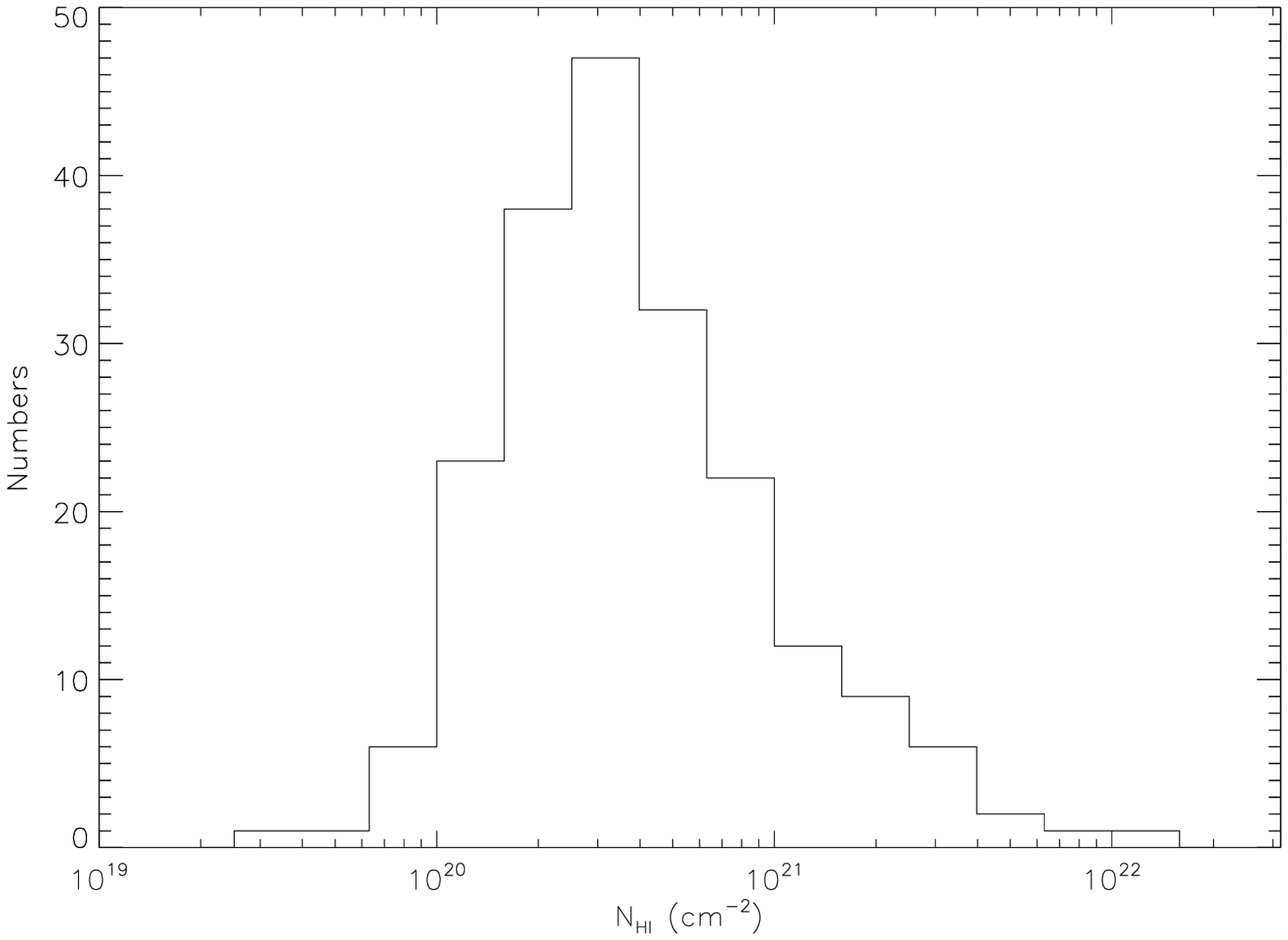}}
  \subfigure[\label{11b}The \hi mass/radius correlation]{\includegraphics[width=7cm]{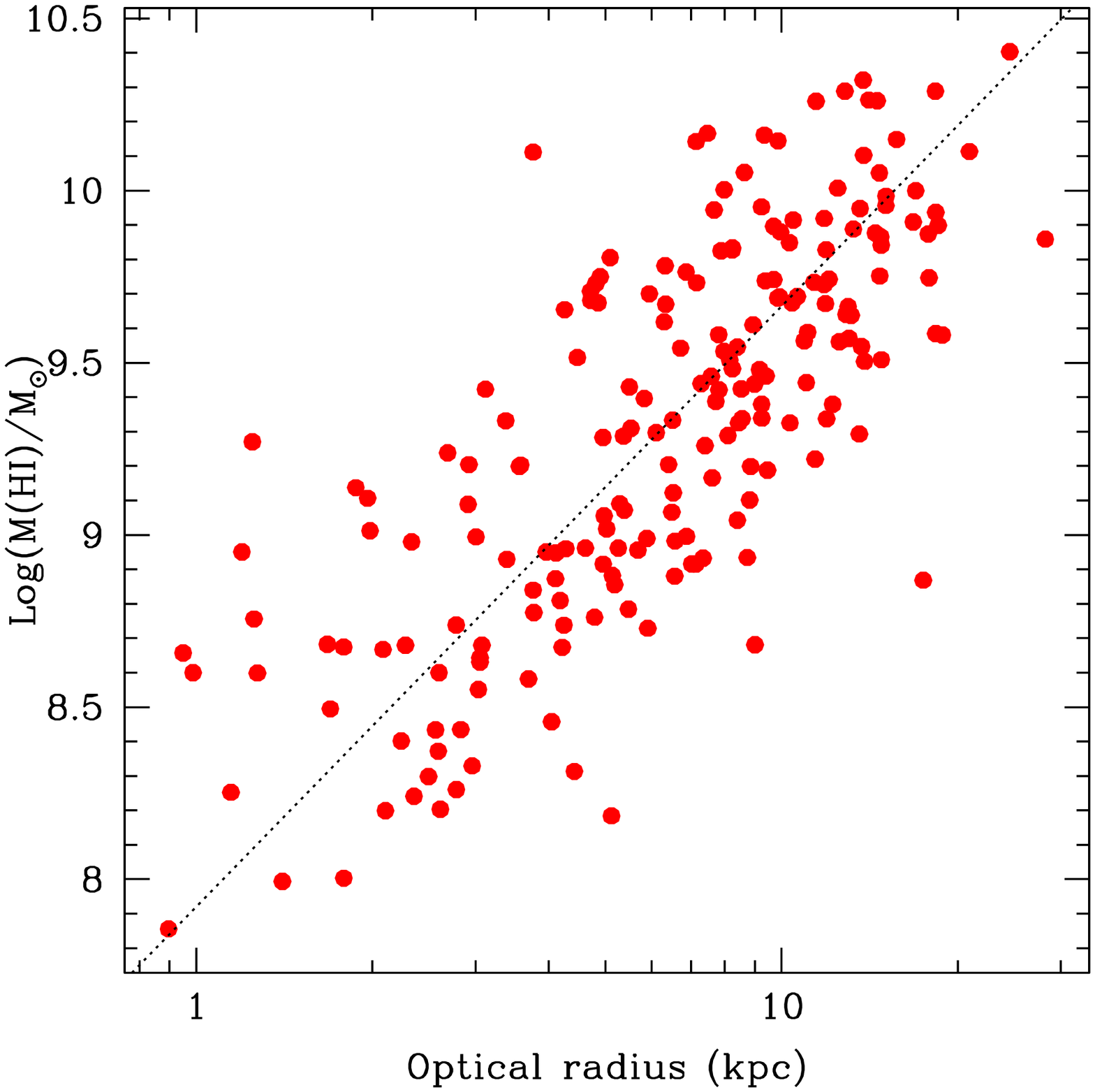}}
  \caption{\label{fig11} (a) shows the Distribution of column
    densities for the ES sample where the densities are obtained by
    assuming $\theta_{21} = 5~\times~\theta_{50}~(opt)$ - see text.
    (b) shows the \hie-mass/radius correlation, i.e. constant column 
density. The dotted line shows the best linear fit to our data.}
\end{figure}

In this section, we show just how diverse the galaxies in a blind \hi survey actually are. 
A quick way to grasp the wide diversity of types within the ES is to examine their SDSS-DR2 
images, which can be seen online (http://www.astro.washington.edu/HIgalaxies) showing a
$g,r,i$ composite image.
They range between red, early type barred spirals several times more luminous than L$_{*}$, containing 
proportionally small amounts of gas [(M$_{H\!I}$/L$_{B}$) $<$ 0.2], and extremely low 
surface brightness objects hundreds of times fainter, very blue, consisting almost 
entirely of gas and so irregular in optical appearance as to be `Inchoate', i.e. 
to be made of nothing more than irregular smudges with apparently unconnected higher 
surface brightness patches here and there. Given just how difficult such Inchoates are 
to see at all on the SDSS-DR2 it is perhaps surprising that not a single source lacks an 
optical counterpart - however ragged it may be. There is even one elliptical galaxy, HIPEQ0154$-$00, 
which appears to have a thin disc and it is being investigated further.

\begin{figure*}
 \begin{center}
 \subfigure[\label{12a} HIPEQ2036-04]{\includegraphics[width=8cm]{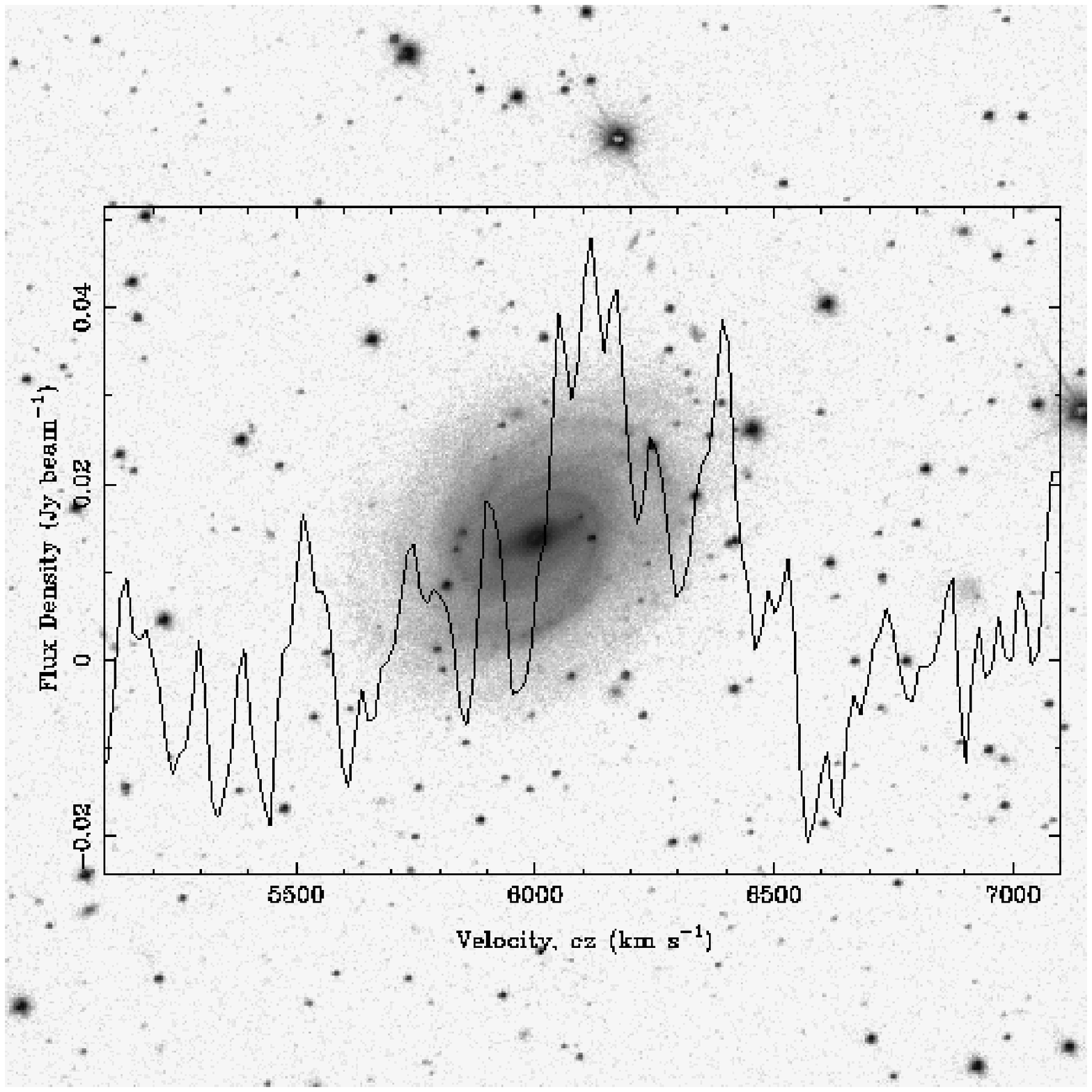}}
 \subfigure[\label{12b} HIPEQ1303+03]{\includegraphics[width=8cm]{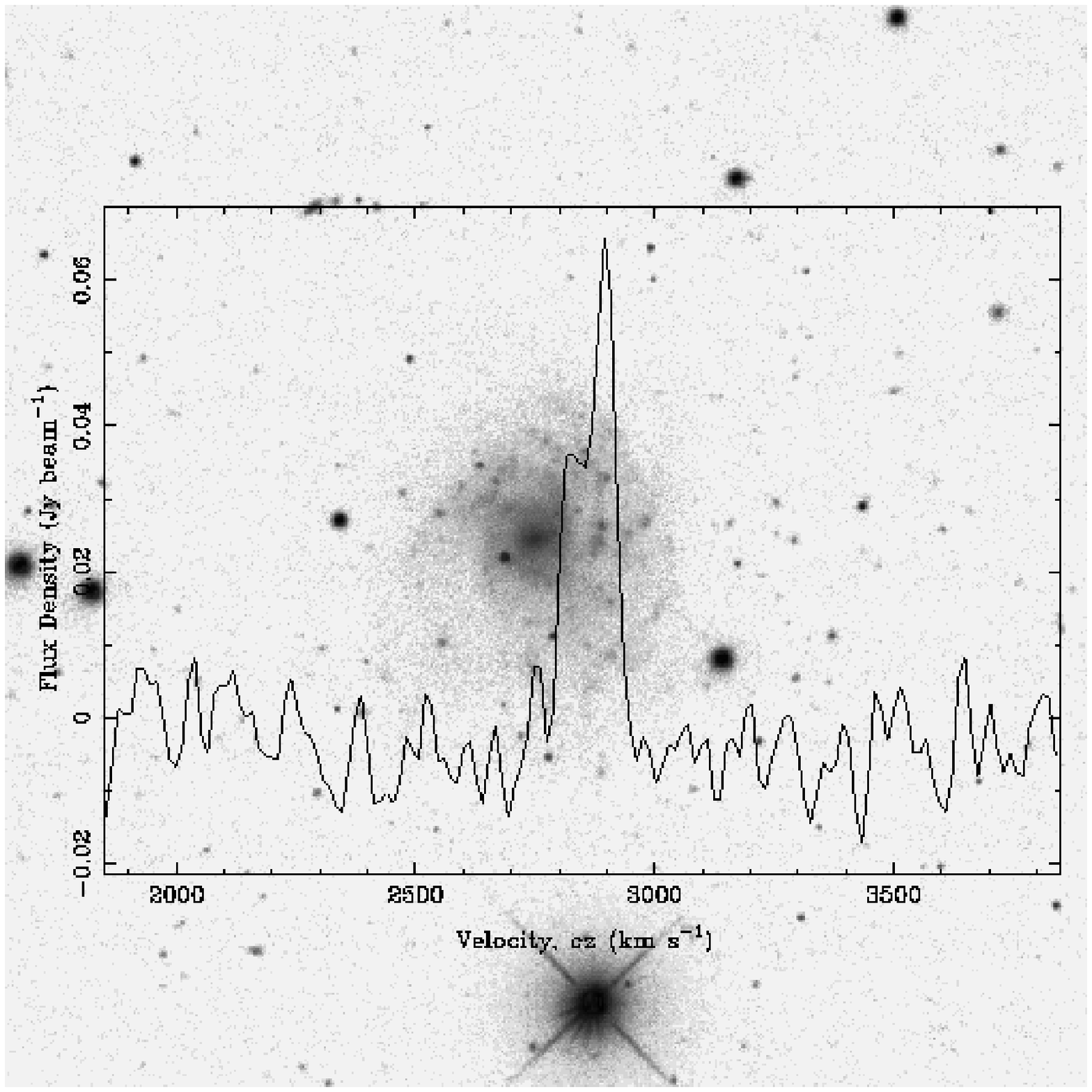}}
 \subfigure[\label{12c} HIPEQ0821-00]{\includegraphics[width=8cm]{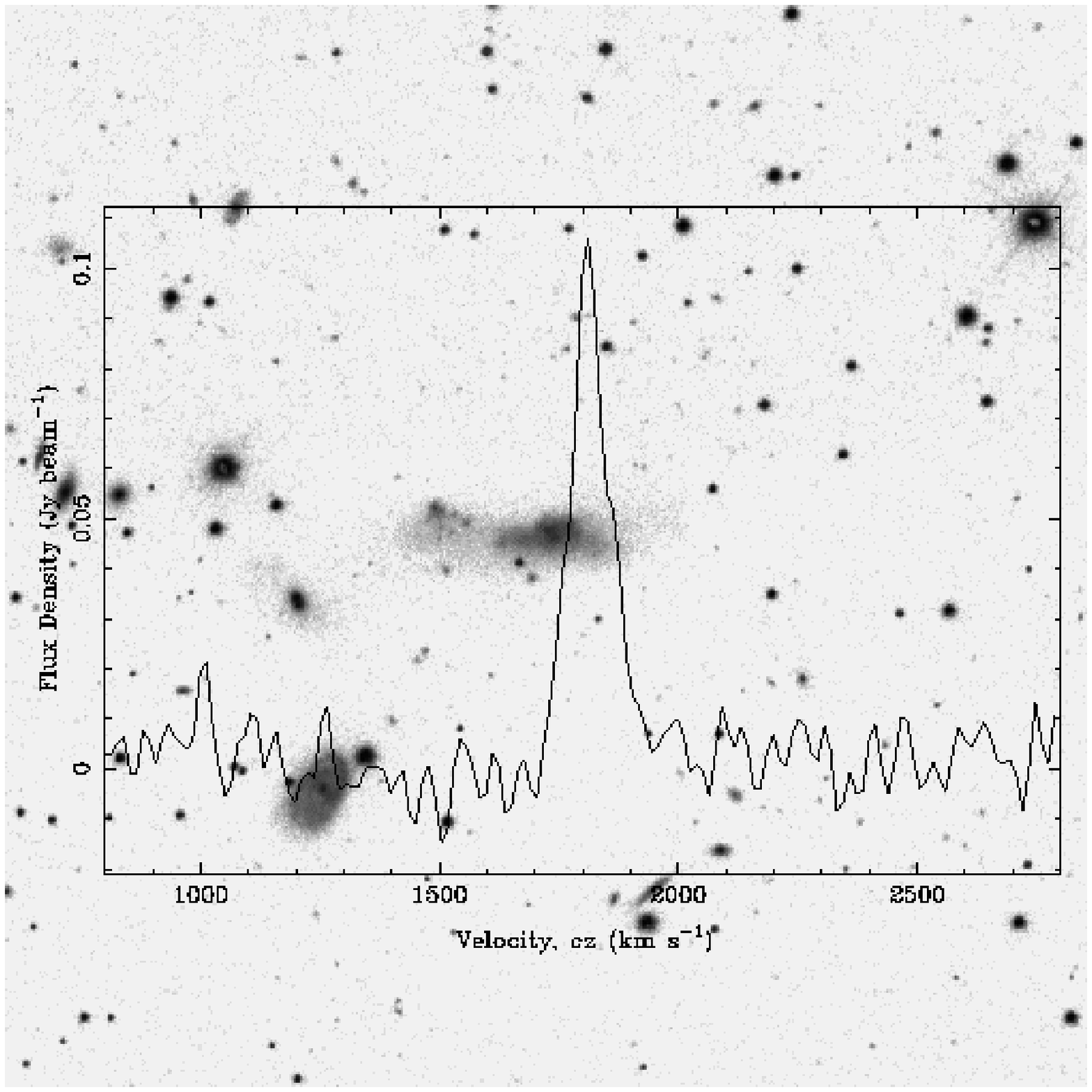}}
  \subfigure[\label{12d} HIPEQ1145+02]{\includegraphics[width=8cm]{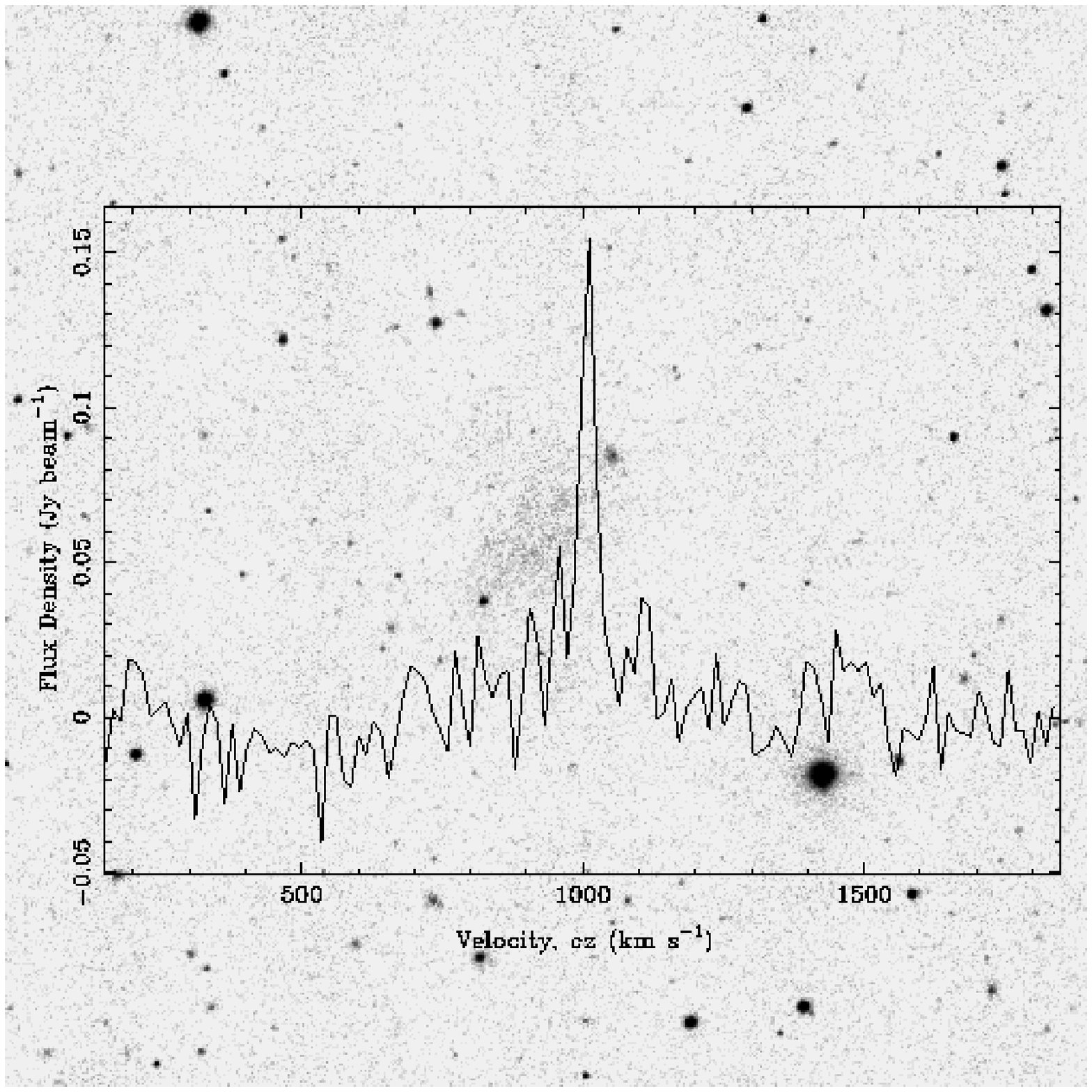}}
 \caption{\label{galimg} These images illustrate characteristic galaxy types within a blind 
\hi survey such as the ES, with their \hi spectra superposed. (a) shows a `Hydrogen Giant'; 
(b) a massive LSBG; (c) an Irregular Galaxy and (d) an `Inchoate Galaxy' - see text.}
  \end{center}
\end{figure*}

To consider the diversity of sources in more detail we consider four individual galaxies 
each representing a different characteristic `class-type' of object that is quite common in the sample. As a 
fiducial comparison one can pick say HIPEQ1507+01 (NGC5850) which is typical of the kind 
of galaxy that turned up in optically selected catalogues and which were afterwards examined 
at 21-cm (e.g. Huchtmeier 1988). It is optically very luminous, 1.7$^{m}$ brighter than 
L$_{*}$, has a low M$_{H\!I}$/L$_{B}$ = 0.15 but nevertheless contains a significant amount 
of \hi (log M$_{H\!I}$ = 9.8) simply because of its high luminosity. It is early type (T=3), 
red [($g-r$) = 0.63] and dynamically very massive (\mbox{log M$_{dyn}$ = 11.7 M$_{\odot}$}) where 
$M_{dyn}=R_{50}(g) \times (W_{20} / sin(i))^{2} / G$. 
 
The greatest number of galaxies in the ES are late type spirals a magnitude or so fainter 
than L$_{*}$ but gas-rich (i.e. median M$_{H\!I}$/L$_{B}$ = 0.91 $\pm$ 0.16) and blue.

\begin{figure}
\centering
\includegraphics[width=7cm]{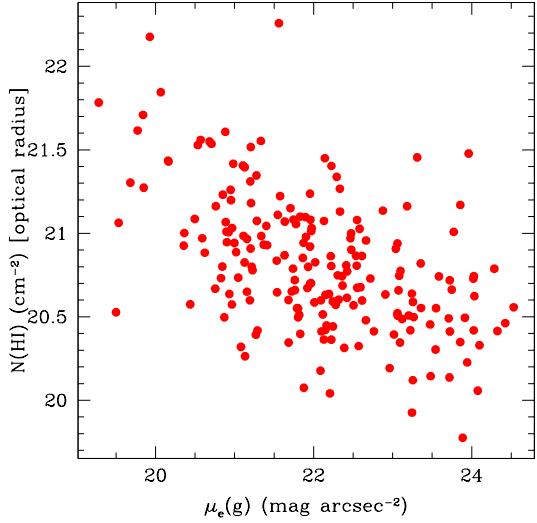}
\caption{\label{SBNHI} The surface brightness/column density
 correlation for the ES galaxies where SB=$\overline{\mu}_{50}(g)$, 
see Table 3.}
\end{figure}

\begin{enumerate}

\item Our first `\hi Selected' type is a \textit{`Hydrogen Giant'}, that is to say a galaxy with more than 
10$^{10}$ M$_{\odot}$ of \hie. HIPEQ2036$-$04 (NGC6941) contains 2 $\times$ 10$^{10}$ M$_{\odot}$, 
apparently an upper limit if one excludes other nearby companions - as here, and about 5 times the \hi 
mass of the Milky Way (Binney \& Merrifield 1998) [Fig \ref{12a}]. Their huge hydrogen content means that such 
galaxies can be detected far away and this one is at 6200 \kmse. It is also a giant optically, being $\sim$ 
3 L$_{*}$ in $g$ but its gas-to-light ratio is M$_{H\!I}$/L$_{B}$ = 0.7 that is 5 times higher than 
the fiducial value for giants. Fig. \ref{12a} shows an SABb (T=3) with a high surface brightness 
core but a low surface-brightness disc with widely spread spiral arms. The integrated colour 
is ($g - r$) = 0.79, while the fibre spectrum of the core is typical of an early type spiral. What is 
remarkable about these Hydrogen Giants is that they appear to have slowed their star-formation 
and so still contain as much mass in gas as in stars. The dynamical mass is high (log M$_{dyn}$ = 11.3), 
but 1.5 times less than NGC5850, the fiducial galaxy. There are 20 in the survey, but given their 
generally high redshifts (median $\sim$ 5500 \kmse) hydrogen giants cannot be common in space. The 
astrophysical challenge is to explain their delayed evolution from gas into stars.

\item Our second class-type is the low surface brightness galaxy of the more massive kind - 
typically an anaemic spiral like HIPEQ1303+03 (UGC08153) shown in Fig. \ref{12b}
(other good examples are HIPEQ1228+02 and 2337+00). It truly is low SB with a 
$\overline{\mu}_{50}(g)$ = 24.3 and yet its dynamical mass is 10$^{10.9}$ M$_{\odot}$. Formally 
speaking there are even more massive LSBGs in the sample but their discs are so 
extremely dim that estimating an accurate axis-ratio, necessary to calculate a dynamical mass, 
often becomes problematic. The total luminosity is 0.5 L$_{*}$ [M$_{g}$ = -18.8] and 
the Hydrogen content is log$_{10}$ M$_{H\!I}$ = 9.0 [M$_{H\!I}$/L$_{B}$ = 0.77]. The colour 
is blue, (g $-$ r) = 0.27, as is commonly the case with LSBGs (McGaugh et al. 1994) but 
there are also some reddish ones in the survey. The question of how significant LSBGs are 
in cosmic terms (e.g. Fukugita et al. 1998) hinges upon two difficult questions: on how massive they 
are and how common. HIPEQ1303+03 and its like confirm, beyond question, that true 
LSBGs can be massive (see also Sprayberry et al. 1993). To what extent blind \hi surveys 
can compensate for the dramatic selection effects against LSBGs in optical catalogues 
is not yet clear. Certainly such surveys find healthy numbers of LSBGs. However Fig. \ref{SBNHI} 
shows that although N$_{H\!I}$ is more or less constant in the sample, there is a perceptible 
fall-off in the column density with SB, with a sharp cut-off at 10$^{19.7}$ \hi atoms cm$^{-2}$. 
This is due to the instrumental limit of the Multibeam system for 400 sec integration so it 
could still be missing some low SB objects (see Minchin et al. 2004). There are $\sim$ 10 LSBGs with 
 dynamical masses $>$ 10$^{10}$ M$_{\odot}$, and as many again which are just as dim apart from a small bright core. 

\item Our third type is the irregular for which HIPEQ0821$-$00 (UGC04358) in Fig. \ref{12c} serves as an 
example though they are very heterogeneous. ES irregulars are naturally gas-rich, HIPEQ0821$-$00 having 
(log M$_{H\!I}$ = 9.3) almost as much \hi as the Milky Way with an M$_{H\!I}$/L$_{B}$ = 3.3. At M$_{g}$ = $-$ 16.4 
it is $\sim$ L$_{*}$/20, has a very blue colour [(g $-$ r) = 0.17] and moderately high SB ($\overline{\mu}_{50}(g)$ = 22.5) 
though SB's among the type vary by 5 magnitudes and global SB's for such irregular objects are rather 
meaningless. There are at least 30 irregulars in the ES sample.

\item The last `type', the `Inchoates', are so dim and faint that they could scarcely be found in any other 
but a blind \hi survey. Our example HIPEQ1145+02 [Fig. \ref{12d}] can barely be seen on the SDSS, having a 
SB $\overline{\mu}_{50}(g)$ = 24.8 at the very limit of the survey. We detect it only because it has a lot of hydrogen for its 
luminosity (M$_{H\!I}$/L$_{B}$ = 7) which at M$_{g}$ = $-$14.31 is 1/200 L$_{*}$. The `Inchoate' label for these 
objects derives from their apparent total lack of organisation. More irregular than irregulars, they have no 
cores or obvious centres, and appear as merely haphazard enhancements of SB at what appear to be HII regions. 
In addition to being extremely gas-rich (M$_{H\!I}$/L$_{B}$ $>$ 5) they are generally blue, thus HIPEQ1145+02, 
by no means extreme in colour, has B $-$ V = 0.48 and (g $-$ r) = 0.27. Other good examples of what is a virtually 
new type of galaxy are HIPEQ0238+00, 0240+01, 0958+01, 1227+01 and 1256+03. We say `virtually' because one of them, 
HIPEQ1227+01, is the famous cloud serendipitously found by Giovanelli and Haynes (1989) and at first thought to be a 
protogalaxy. Indeed it would have been the easiest Inchoate to find as it has the highest M$_{H\!I}$/L$_{B}$ ($\sim22$) 
of \textit{any} object in the ES sample.
\end{enumerate}

Now that we have a dozen or so `Inchoates' in the ES to study, some of the puzzling questions raised by the original 
Giovanelli and Haynes cloud return with even greater insistence. It is the \textit{combination} of their properties which 
makes it difficult to explain Inchoates (Salzer et al. 1991, Grossi et al. 2007). Their 
extraordinarily high gas-mass fraction indicates little integrated past star formation, while their blue colours 
[as blue as (B$-$V) $\leq$ 0.3] can only be explained with star formation that rises sharply to the present day 
(Bruzual and Charlot 2003). However, if such a rise is only a temporary burst, then the galaxies should soon fade by 
1.5 magnitudes while reddening from (B$-$V) = 0.3 to 0.5, leaving behind an optically undetectable dark \hi cloud 
(Leitherer et al. 1999). However there are no such dark clouds in our survey, and none among 4315 HIPASS sources 
detected across half the sky (Doyle et al. 2005), all of which are optically detected in either SDSS or the SuperCosmos 
Sky Survey (Hambly et al. 2001). Thus a bursting explanation for their blue colours seems less likely than either 
a truly young galaxy or a steadily rising star formation rate. However, GALEX measurements of 2 of them (1145+02 and 
1256+03) imply current star formation rates that are only a factor of $\sim$ 2 greater than the past average. 
And if they were truly young, why do not blind \hi surveys like HIPASS find their predecessor dark protoclouds? `Inchoates' 
are indeed a puzzle.

Whatever they are, Inchoates make a dramatic contrast with fiducial galaxies like NGC5850 and with the higher SB spirals 
which make up more than half the sources in the ES sample. Together with luminous gas-poor galaxies, hydrogen giants, 
massive LSBGs and irregulars they illustrate a diversity which suggests a wide scatter among several underlying fundamental 
parameters such as mass, age, angular momentum and binding energy. But such turns out not to be the case.

\begin{figure}
\centering
\subfigure[\label{14a} $(r-i)$~vs~$(u-g)$ relation]{\includegraphics[width=7cm]{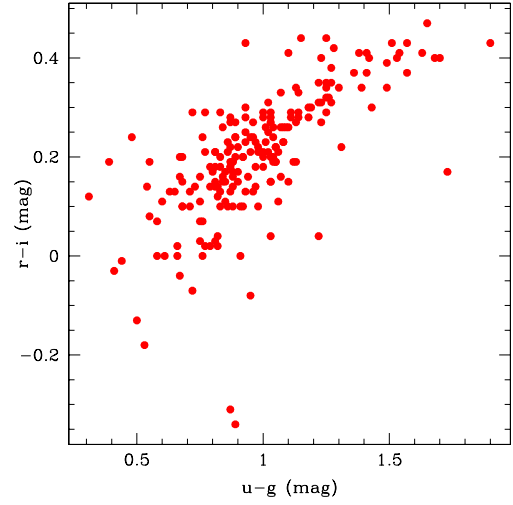}}
\subfigure[\label{14b} $(i-z)$~vs~$(g-r)$ relation]{\includegraphics[width=7cm]{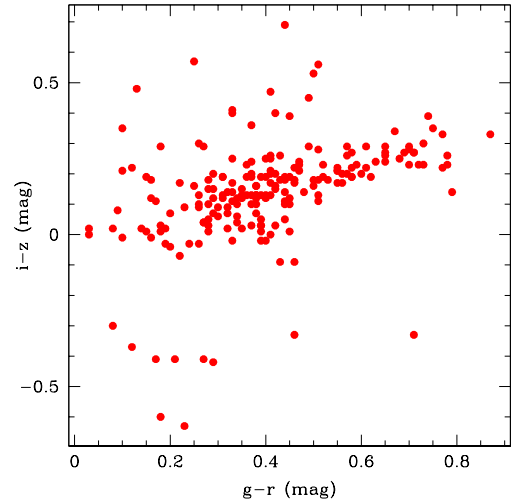}}
\caption{\label{colours} (a) shows the relations between the $(u-g)$
  and $(r-i)$ optical colours and (b) shows the relation between the
  $(g-r)$ and $(i-z)$ optical colours of the ES sample.}
\end{figure}

\section{CORRELATIONS AMONG GALAXY PROPERTIES}

We have measured 9 quantities in all, combined with the inclination (b/a) which can be 
used to correct observed quantities to face on: mass of \hie, \hi line width (as defined in Section 2), 
luminosity (in $g$), four colours ($u-g$), ($g-r$), ($r-i$), and ($i-z$), and 2 radii (R$_{50}$ and R$_{90}$ in $g$ band) 
all defined in Section 3.
What we are chiefly missing is some kind of environmental parameter, which could be vital. 
However we have been so fastidious in excluding closely packed galaxies that we cannot use 
environment for now. Later, with more interferometry of the dubious identifications within groups, 
we can add that to the study. Even so it is worth pressing ahead without it, provided we remember 
the caveat, because Gavazzi et al. (1996) showed that for the late type galaxies which appear in the 
ES, environment appears to be unimportant.

\begin{figure}
\centering
\includegraphics[width=7cm]{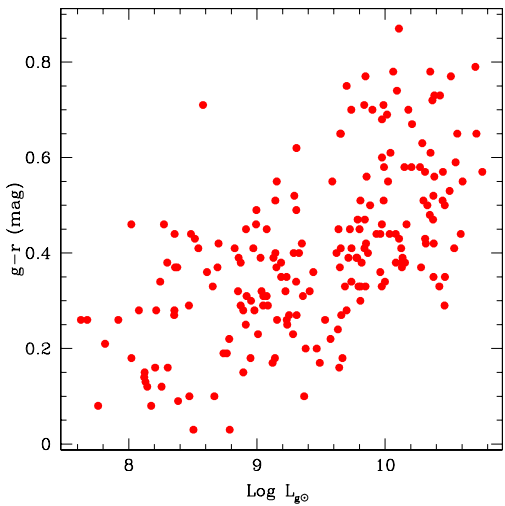}
\caption{\label{colourmag} The ($g-r$) colour/luminosity relation.}
\end{figure}

From the 9 basic measurements, and the inclination and assumed distance, one can construct other 
more familiar and perhaps more interesting parameters such as circular velocity ($V_{0}$), 
dynamical mass $M_{dyn}$ ($\sim$~$V_{0}^{2}R_{50}/G$), surface brightness $\overline{\mu}_{50}(g)$  ($\sim L_{g}/R_{50}^{2}$), 
column density ($N_{H\!I} \sim M(H\!I)/R_{90}^{2}$), brightness temperature T$_{H\!I}$ ($\sim N_{H\!I}/\Delta W_{20}$), 
angular momentum per unit mass ($q\sim V_{0} R_{90}$) and so on. With at least 15 such synthetic parameters 
to work with there are $\approx$ (15$\times$14/2) $\approx$ 100 correlations to look for. 
We could have used some statistical technique but we elected to look at all the correlations by eye for now.
Correlation coefficients, and their significance are given for each correlation. 
Thus one can discount obvious selection effects, and the already well-known correlations, to concentrate 
on looking for what might be new or surprising.

\begin{figure}
\centering
\includegraphics[width=7cm]{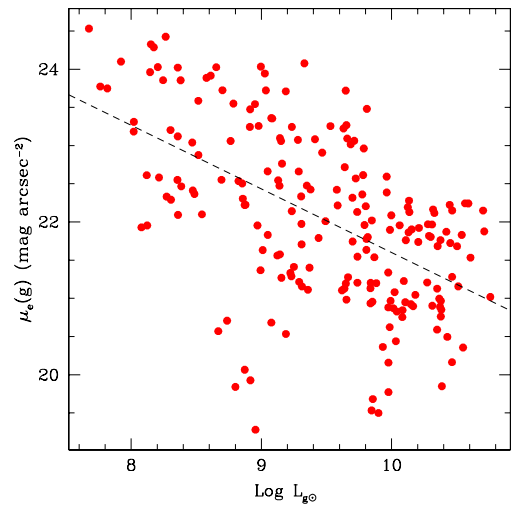}
\caption{\label{SBlum} The surface brightness/luminosity 
relation in $g$ band. The dotted line shows the best linear 
fit to the data.}
\end{figure}

\begin{figure}
\centering
\includegraphics[width=7cm]{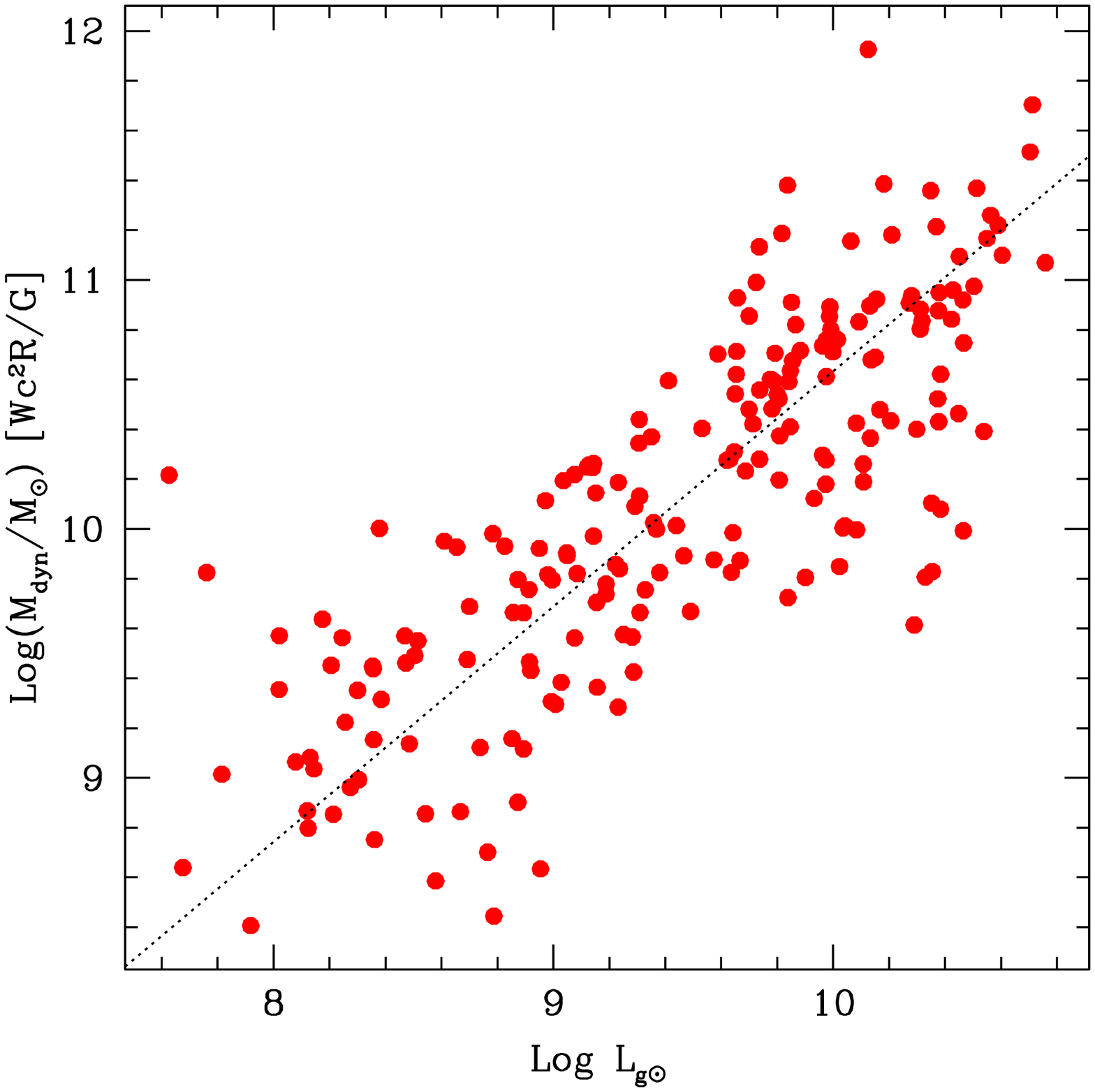}
\caption{\label{dynlum} The dynamical mass/luminosity relation 
in $g$ band. The dotted line shows the best linear fit to the data.}
\end{figure}

We are of course far from the first to attempt such a systematic 
search among galaxy properties. For instance Brosche (1973), 
Balkowski (1973), Tully \& Fisher (1981), Gavazzi et al. (1996), 
Blanton et al. (2003) looked at optically selected samples but with 
added \hi data, whilst Minchin et al. (2003), Rosenberg et al. (2005), 
Kovac (2007) and Begum et al. (2006) have recently looked at \hi
selected sets followed up either by optical or by NIR measurements
from 2MASS. Elsewhere we shall compare results, but for now we examine 
only our own correlations with an eye unbiased by previous work.

As a start it is important, given the rather small number in the sample (195), 
to establish that there is sufficient dynamic-range within its intrinsic properties to make correlation 
analyses worth-while. To look for relations between two galaxy properties $\textit{y}$ and $\textit{x}$, of the 
form $y = Ax^{m}$, it is elementary to show that the error on \textit{m} due to measurement errors $\delta y$, $\delta x$, as 
opposed to real scatter in the properties themselves, is given by:

\begin{equation}
\sqrt{\left\langle \frac{\delta m}{m} \right\rangle^{2}} \approx 1.5 \sqrt{\left( \frac{(\delta y/y)}{\Delta(mag y)}\right)^{2} + \left( \frac{(\delta x/x)}{\Delta (mag x)} \right)^{2}}
\end{equation}

where $magx \equiv 2.5~log_{10}(x)$ and the $\Delta (mag)$'s 
give the relevant dynamic ranges. 
Table \ref{ranges} shows in \textit{Column (1)} the most relevant properties measured for the
galaxies in the ES sample.\\
\indent \textit{Column (2)}.-- Median value of the measurement.\\
\indent \textit{Columns (3) and (4)}.-- The maxima and minima values
respectively, where the few objects in the extreme wings of 
the distribution have been ignored.\\
\indent \textit{Column (5)}.-- Dynamic range in magnitudes.\\
\indent \textit{Column (6)}.-- Shows the estimated relative measurement errors, 
$1.5 (\delta x/x)/\Delta (mag~x)$.\\
In most cases the relevant measurement errors are small enough 
($<$ 0.1) not to dominate a correlation with another quantity 
in the table, the main exceptions being line-width $W_{20}$ and 
colour. We do not use line-width except in the combination 
$W_{20}^{2}R/G$, i.e. dynamical mass, where the relative 
measurement errors are reduced by the large dynamic range to 
\mbox{$<$ 0.06}.
\begin{table*}
\caption{Properties of the ES sample with optical SDSS-DR2 data.}
\label{ranges}
\begin{center}
\begin{tabular}{lccccc}
\hline
\multicolumn{1}{c}{\textbf{Property}} & 
\multicolumn{1}{c}{\textbf{Median}} & 
\multicolumn{1}{c}{\textbf{Min}} & 
\multicolumn{1}{c}{\textbf{Max}} & 
\multicolumn{1}{c}{\textbf{Range}} & 
\multicolumn{1}{c}{\textbf{Relative error}}\\
\multicolumn{1}{c}{(1)} &
\multicolumn{1}{c}{(2)} &
\multicolumn{1}{c}{(3)} &
\multicolumn{1}{c}{(4)} &
\multicolumn{1}{c}{(5)} &
\multicolumn{1}{c}{(6)} \\
\hline
Mass \hi ($M_{\odot}$) & 10$^{9.38 \pm 0.40}$ & 10$^{7.5}$ & 10$^{10.2}$ & 6.8 & .03 \\
Luminosity ($g$) ($L_{\odot}$) & 10$^{9.51 \pm 0.38}$ & 10$^{7.2}$ & 10$^{10.6}$ & 8.5 & .01 \\
Area ($\pi R_{50}^{2}$) (kpc$^{2}$) & 10$^{1.91 \pm 0.1}$ & 10$^{-0.2}$ & 10$^{3.12}$ & 8.3 & .01\\
Line-width W$_{20}$ (\kmse) & 10$^{2.12 \pm 1.78}$ & 10$^{1.60}$ & 10$^{2.60}$ & 2.5 & .09\\
Surface brightness ($L_{\odot} pc^{-2}$) & 10$^{1.85 \pm 0.07}$ & 10$^{0.88}$ & 10$^{2.99}$ & 5.2 & .01\\
Dynamical mass ($M_{\odot}$) & 10$^{10.14 \pm 0.53}$ & 10$^{8.5}$ & 10$^{11.3}$ & 7.0 & .06\\
Colour ($g - r$) & 0.39 $\pm$ 0.11 & 0.1 & 0.7 & 0.6 & .07\\
\hline
\end{tabular}
\end{center}
\end{table*}

First notice that the 4 colours are strongly correlated (Fig. \ref{colours}), as was to be expected, so there is considerably 
degeneracy among them and, as revealed already, to first order there is only 1 independent colour (Strateva et al. 2001).
Many properties correlate with luminosity and Fig. \ref{colourmag} shows 
the colour-luminosity diagram, which is well known for optically 
selected galaxies, i.e. more luminous galaxies are redder. When it 
comes to disentangling the physics going on the actual slopes in all 
such correlation diagrams will be crucial. Thus Fig. \ref{colourmag} 
implies that L$_{r}$ $\sim$ L$_{g}^{1.1}$ roughly. The Pearson correlation 
coefficient is $r_{p} \sim$ 0.61, corresponding to a probability $P(r>r_{p})~>99.9\%$. 
This is the first of the 5 correlations.

Fig. \ref{SBlum} exhibits the second correlation ($r_{p} \sim$ $-$0.55, $P(r>r_{p})~>99.9\%$), 
i.e. between surface-brightness and luminosity and, not unexpectedly less luminous galaxies are dimmer. 
While this has been long suspected (e.g. Bingelli 1984) one could never be certain, 
in any optically selected sample, to what extent it was an artefact of the 
dramatic selection effects acting on surface-brightnesses which lie so close to 
the sky (e.g. Disney 1999). However here all the galaxies have been identified without 
regard to their optical properties and it seems very likely that Fig. \ref{SBlum} 
tells us something fundamental about the relation between luminosity and radius with few selection effects. While 
the scatter is large the correlation is clear and very roughly speaking the surface 
brightness $\Sigma(g) \sim L_{g}^{0.5}$.

The third correlation ($r_{p} \sim$ 0.75, $P(r>r_{p})~>99.9\%$), between \hi mass and optical radius, is 
illustrated in Fig. \ref{fig11}. It is a tight one and suggest, that all galaxies have 
the same \hi column density, i.e. $M_{H\!I} \propto R_{g}^{2}$. This was discovered 
first by Haynes and Giovanelli (1984) in an optically selected sample,
discovered again in an \hi selected sample by Minchin et al. (2003)
and confirmed in an interferometer study by Rosenberg et al. (2005) 
in the Arecibo Dual Beam Survey and Kovac (2007) and our ES measurements concurs.
For now constant surface density, is an intriguing puzzle which needs to be explained.

\begin{figure}
\centering
\includegraphics[width=7cm]{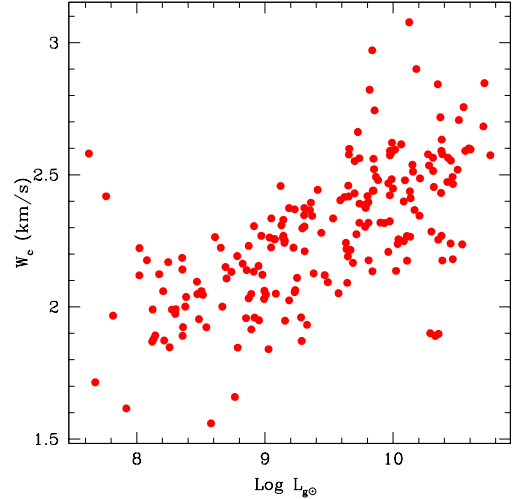}
\caption{\label{TFg} The Tully-Fisher relation in $g$ band.}
\end{figure}

\begin{figure}
\centering
\includegraphics[width=7cm]{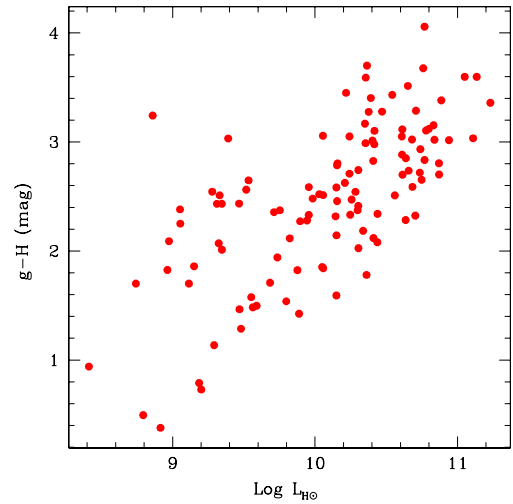}
\caption{\label{Hbandlum} The ($g - H$) colour H band luminosity relation for the ES galaxies detected by 2MASS.}
\end{figure}

\begin{figure}
\centering
\includegraphics[width=7cm]{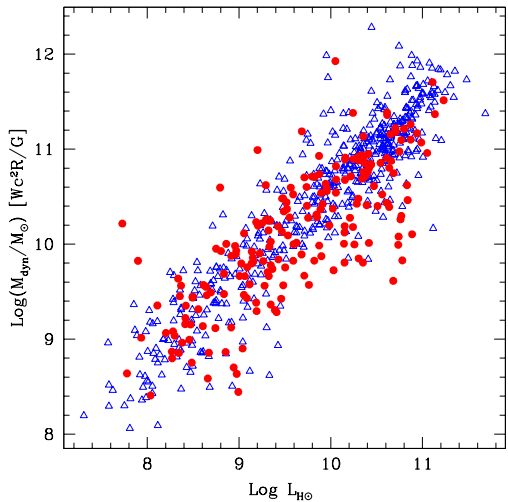}
\caption{\label{gavHGM} The dynamical mass-H band luminosity relation for our samples (circles) and for the 
GOLDMine sample (triangles).}
\end{figure}

The fourth correlation ($r_{p} \sim$ 0.81, $P(r>r_{p})~>99.9\%$) we find is between luminosity ($g$) and 
dynamical mass shown in Fig. \ref{dynlum}. Over more than 3 orders of magnitude the $g$ luminosity is tightly 
correlated with dynamical mass, a correlation that appears perceptibly tighter than the 
Tully-Fisher correlation which we show in Fig. \ref{TFg} (see later). The slope is close to 1, implying a direct 
proportion between dynamical mass and luminosity. This is not new either but it was 
discovered in the remarkable paper by Gavazzi et al. in 1996. In order to compare our data
 with Gavazzi's, we transform ours into the $H$-band, which should be freer of dust and of 
 transient star-formation effects. 113 of our ES sources are in 2MASS and Fig. \ref{Hbandlum} 
 illustrates the relation between their colour and their luminosity, which implies:
\begin{equation}
\label{GavEq}
\log(L_{H}) = (1.31 \pm 0.05) \times log(L_{g}) - (2.9 \pm 0.4)
\end{equation}
i.e. $L_{H} \sim L_{g}^{4/3}$.
 If we use this to transform our Fig. \ref{dynlum} data into the H-band we reach Fig. \ref{gavHGM} 
 where our galaxies are plotted on top of the GOLDMine sample assembled by Gavazzi and colleagues 
 (2003) for a mixture of bright discs and Virgo dwarfs.
 Over almost 4 orders of magnitude, disc-like galaxies have $M_{dyn} \sim L_{H}$ with a scatter 
 of only $\sim$ 1 mag about the relation in both axis directions. Whether they are selected optically 
 (Gavazzi), or by their neutral hydrogen signal as here, the 635 galaxies all appear to have the 
 same $H$-band Mass-to-Light Ratio $\sim$ 4 in solar units. 
 The numerical value 4 is an artefact of using $R_{50}(g)$ in the conventional definition of $M_{dyn}$, which 
 we have so far adopted. However $M_{dyn} \propto R\times(W_{20}sin(i))^{2}$ and as the outermost HI 
 test particles are typically at $\sim$ 5 $R_{50}(g)$ in images 21-cm galaxies (Salpeter \& Hoffman 1996, Kovac 2007) 
 a more accurate $M_{dyn}$ would be $\sim$ 5 times greater than we and Gavazzi have implied. In 
 that case the $H$-band mass-to-light ratio for both our sample, and the GOLDMine sample, is $\sim$ 20, leaving 
 plenty of scope for dark matter. Note that this ratio cannot be luminosity dependent (see Fig. 20).
  `Gavazzi's relation' as we call it, is a remarkably uniform one for any \mbox{extragalactic} data-set.
 
\begin{figure}
\centering
\subfigure[\label{21a}]{\includegraphics[width=7cm]{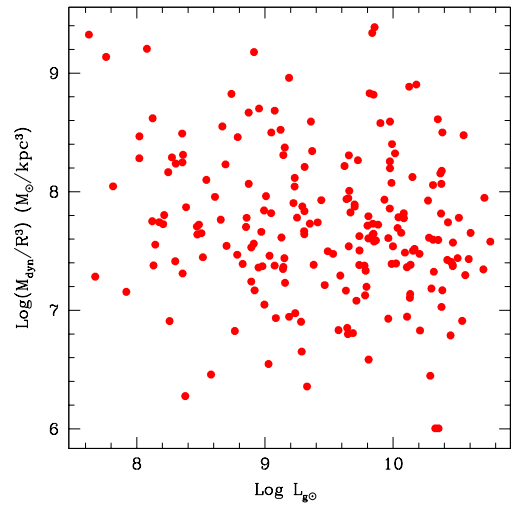}}
\subfigure[\label{21b}]{\includegraphics[width=7cm]{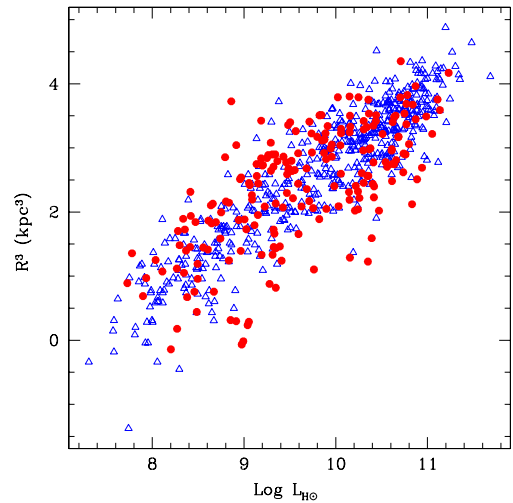}}
\caption{\label{MR3} Top: The ratio $M_{dyn}/R^{3}$ ratio as a function of the $g$ band luminosity. Bottom: The 
relation between the dynamical mass and $R^{3}$ for our sample (circles) and for the GOLDMine sample (triangles).}
\end{figure}

\begin{figure}
\centering
\includegraphics[width=7cm]{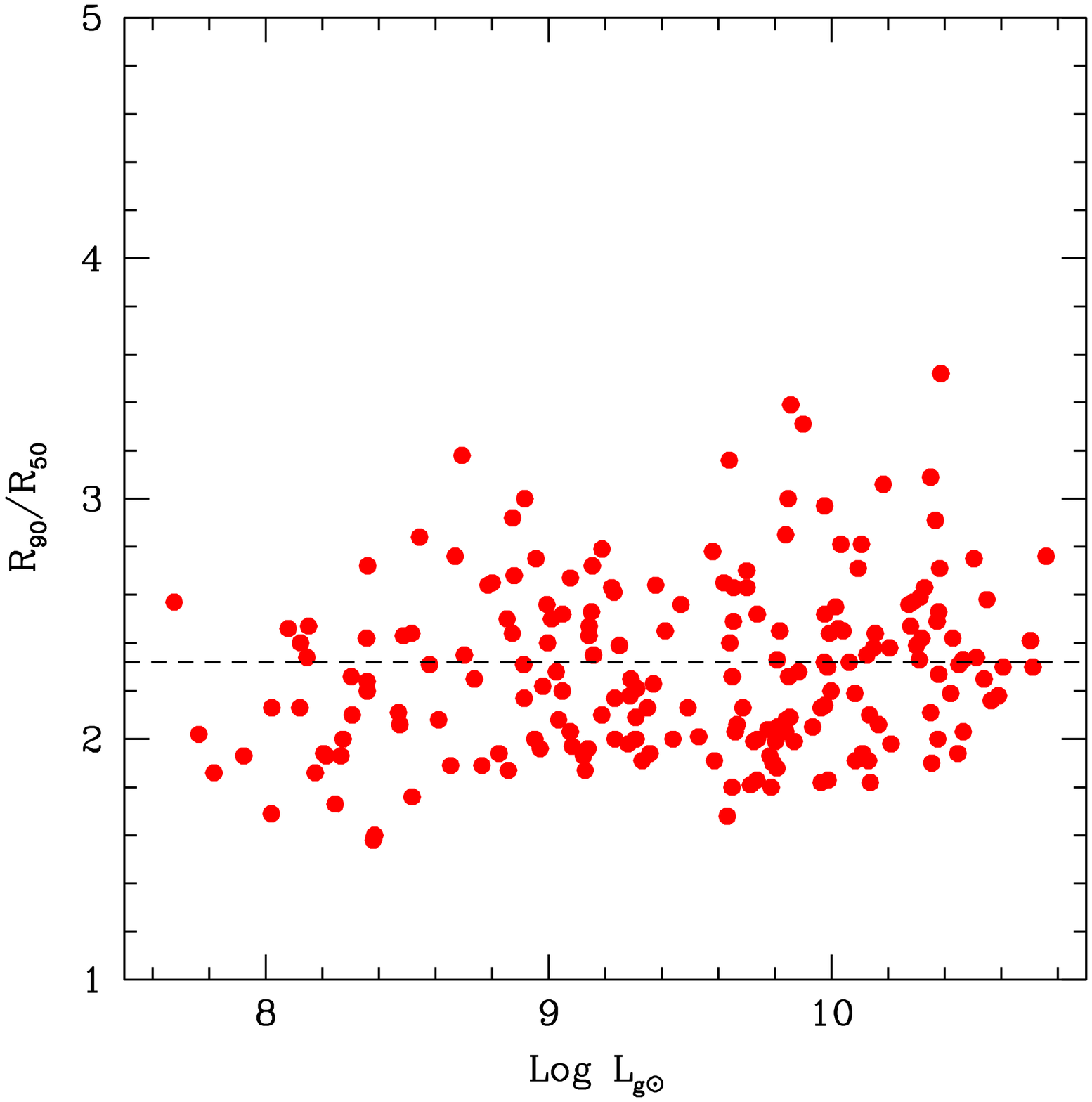}
\caption{\label{R90R50} The concentration index of \hi galaxies as a function of luminosity. A best fit line has a slope 
of only -0.017. The ratio should be 2.32 for a perfect exponential disk. }
\end{figure}

The fifth and last correlation (r$_{p}$ $\sim$ 0.18, $P(r>r_{p}) \sim$ 99.9$\%$) we look at in Fig. \ref{R90R50} is that 
between the two separate radii measured by SDSS i.e, $R_{50}$ and $R_{90}$. The very tight correlation, and the ratio 
between them of 2.3 means that despite the very wide range of galaxy types in the ES their outer light 
profiles are all well fitted by an exponential law and the ratio does not change with luminosity.

\section{DISCUSSION}

We set out to look for correlations within a sample of galaxies free of optical 
selection effects. Apart from inclination and velocity we measured 9 parameters 
for each galaxy. Since the 4 colours appear degenerate we are left with 6 
$\textit{a priori}$ independent observables, between which we find the 
following 5 correlations, 4 old and 1 new (A) which are shown in Table 
\ref{correlations}, where the columns are as follows.\\
\indent \textit{Column (1)}.-- Correlations observed.\\
\indent \textit{Columns (2) and (3)}.-- Denote the distance dependance, d$^{x}$, 
of the left and right-handed parameters in the correlation.\\
\indent \textit{Column (4)}.-- Indicates the difference between 
the two exponents in Cols. (2) and (3) (see text for explanation).\\
\indent \textit{Column (5)}.-- Shows the $\textit{observables}$ 
introduced into the correlation, that have not been used before.

\begin{table}
\caption{Correlations in the ES sample.}
\label{correlations}
\begin{center}
\begin{tabular}{lccccl}
\hline
 & \multicolumn{1}{c}{(1)} &
\multicolumn{1}{c}{(2)} &
\multicolumn{1}{c}{(3)} &
\multicolumn{1}{c}{(4)} &
\multicolumn{1}{c}{(5)} \\
\hline
\multicolumn{1}{c}{A} &
\multicolumn{1}{l}{SB/Lum} &
\multicolumn{1}{c}{d$^{0}$} &
\multicolumn{1}{c}{d$^{2}$} &
\multicolumn{1}{c}{2} &
\multicolumn{1}{l}{SB, $\theta_{90}$, d}\\
\multicolumn{1}{c}{B} &
\multicolumn{1}{l}{Col/Lum} &
\multicolumn{1}{c}{d$^{0}$} &
\multicolumn{1}{c}{d$^{2}$} &
\multicolumn{1}{c}{2} &
\multicolumn{1}{l}{$(g - r)$}\\
\multicolumn{1}{c}{C} &
\multicolumn{1}{l}{M$_{dyn}$/L$_{H}$} &
\multicolumn{1}{c}{d$^{1}$} &
\multicolumn{1}{c}{d$^{2}$} &
\multicolumn{1}{c}{1} &
\multicolumn{1}{l}{$\Delta V$}\\
\multicolumn{1}{c}{D} &
\multicolumn{1}{l}{R$_{90}$/R$_{50}$} &
\multicolumn{1}{c}{d$^{1}$} &
\multicolumn{1}{c}{d$^{1}$} &
\multicolumn{1}{c}{0} &
\multicolumn{1}{l}{$\theta_{50}$} \\
\multicolumn{1}{c}{E} &
\multicolumn{1}{l}{M$_{H\!I}$/R$_{90}^{2}$} &
\multicolumn{1}{c}{d$^{2}$} &
\multicolumn{1}{c}{d$^{2}$} &
\multicolumn{1}{c}{0} &
\multicolumn{1}{l}{S~(21cm)}\\
\hline
\end{tabular}
\end{center}
\end{table}


Column (5) in Table \ref{correlations} is pertinent to the 
$\textit{independence}$ of the 5 correlations. Unless $\textit{a priori}$ 
reasons can be advanced for a dependance of the new observables on previous 
ones, then the 5 correlations must be independent. It might, for instance, 
be argued that $\theta_{50}$ should depend on $\theta_{90}$, but that 
ignores the real differences in `concentration' q $\equiv$ 
$\theta_{90}$/$\theta_{50}$ between galaxies of different 
morphological type (e.g. q = 2.3 for pure exponential but q = 4.2 for 
de Vaucouleur profiles). Concentration is a crude but quantitative 
proxy for morphological type, and there is no $\textit{a priori}$ case
for this to be the same among HI-sources. Thus we can argue that the 5 
correlations are truly independent.

Although the number of galaxies is modest the dynamic range of most of their properties is large 
(i.e. more than 7 magnitudes in luminosity, area, M$_{H\!I}$ and M$_{dyn}$) whilst the correlations are, in 
a statistical sense, highly significant. So far as the $\textit{existence}$ of the correlations (our main interest here) 
is concerned the sample-size is not an embarrassment, though larger samples will be needed to pin down the $\textit{slopes}$ 
of the correlations more precisely. There is moreover a great diversity of galaxy types in the sample, ranging 
between early-type giant spirals and tiny `Inchoates' of such low
surface brightness that they would usually be, and 
indeed were, missed in optical surveys. The difficulty of making a reliable census of galaxies in light 
alone is illustrated by Fig. \ref{SBlum} where it can be seen that at all luminosities galaxies can vary 
in SB over 3 or 4 magnitudes, making the dimmest of them, even the giants, hard to pick out from the 
background by optical means alone. The narrow SB range of earlier catalogues (Freeman 1970, Disney 1976) was 
almost certainly a selection artefact.

Two correlations in particular need explaining. Correlation (E), 
i.e. M$_{H\!I}$ $\sim$ R$^{2}_{g}$ could be related to the stability of 
a gas layer to gravitational collapse, and hence to a trigger for star formation 
(e.g. Kennicutt 1989, Schaye 2008). The other, the SB/Luminosity correlation (A) 
has considerable scatter (Fig. \ref{SBlum}) making it difficult to 
pin down an accurate value for $\beta$ in $\Sigma \sim L^{\beta}$, 
where $\beta$ is not far from 1/2.
Combining our data with the GOLDmine sample, in order to increase 
sample-size and hopefully reduce scatter, yields $\Sigma \sim 
L^{1/3}$, or $\Sigma \sim R$, or $L_{H}/R_{g}^{3}$ = constant or, 
considering Gavazzi's Law, M$_{dyn}$/R$_{g}^{3}$ = constant ($\sim$ 1 
H atom cm$^{-3}$ within the optical radius) which, see Fig. 
\ref{21b}, is a pretty good fit. So the mass - and luminosity - 
density of galaxies is more or less independent of their Luminosity. 
This is at least a better mnemonic for recalling the rather scattered 
SB/Lum law. If true the Virial Theorem implies that all HI galaxies 
spin at the same angular velocity and rotate $\sim$ 40 times in a 
Hubble epoch.

What about the Tully-Fisher correlation $\Delta V \sim L^{\alpha}$? 
Gavazzi's Law (C): $L_{H} \sim M_{dyn} \equiv R_{g} \Delta V^{2}$ 
while above: $R_{g} \sim L_{H}^{1/3}$, hence $\Delta V \sim
L_{H}^{1/3}$. In other words our data are consistent with the TF law provided 
$\alpha$ = 1/3 [see also Fig. \ref{TFg}]. The problem with using 
TF directly is demonstrated in Table \ref{ranges} where it can be 
seen that the low dynamic range in $\Delta V$ will always make it 
difficult to find the exponent $\alpha$ accurately, and it makes 
more sense to incorporate $\Delta V$ into the Dynamical Mass 
($\equiv$ R $\Delta V^{2}$) instead, as it is in Gavazzi's Law (C) 
where it leads to both a clear correlation with Luminosity 
(Fig. \ref{gavHGM}) and perhaps a natural physical explanation 
[`all galaxies are made of the same stuff'].

The distance to most galaxies in our radial velocity range can be problematic so it is important to ask if and how 
distance uncertainties will affect the correlations. Where there is a difference (last column) between the 
distance-dependant exponents $x$ in our list one expects distance uncertainties to introduce extra scattering to 
correlations, as for instance in (A) but not in (E). It is interesting to note that the tightest correlations 
observed i.e. (D) and (E) are the ones unaffected by distance uncertainties, whilst the loosest, i.e. (A) 
[Fig. \ref{colourmag}] and (B) [Fig. \ref{SBlum}] should indeed be the strongest affected. This hints that 
distance uncertainties probably are increasing the scatter. On the other hand look at Gavazzi's relation (B) 
and in particular Figures \ref{dynlum} and \ref{gavHGM}. There is no increase in scatter when we come to 
lower radial velocities where one might expect the relative distance-uncertainties to have the most pernicious 
effect on the correlations. This suggests that although distance uncertainty can increase scatter, that increase 
must be rather limited here, and exculpates the uniform but rather crude distances we have adopted. Note also 
that only 1 observable is size, and hence distance dependant (R 
$\equiv d \theta_{90}$) while 5 are not, i.e. $\Sigma_{opt}, \nhi, 
(g - r), \Delta V$ and Concentration $R_{90}/R_{50}$. This implies 
strong correlations among $\textit{intrinsic}$ galaxy properties 
(e.g. colour and SB, $\Delta V$ and \nhi) i.e. ones that have nothing 
to do with either scale or distance.

How constraining could the five discussed correlations (A) to (E) be on theories of galaxy formation and evolution? 
That depends on how many fundamental, independent invariant physical properties galaxies might posses. At present 
it is hard to imagine more than 7: total mass, baryon fraction, age, angular momentum, random energy, radius and 
central condensation. Once a galaxy has virialised, the Virial Theorem will provide one relation between them, 
leaving only 6 independent properties. Ignoring interactions one might expect, to first order, that these 
properties might be conserved by each galaxy though, on a secular time-scale some random energy might be dissipated, 
while some baryons might be ejected. Thus five correlations (if they are truly independent) within a six-parameter 
set hint at a high degree of organisation among gas-rich galaxies, something partially foreshadowed by the early 
\hi pioneers [Brosche 1973, Balkowski 1973]. We are presently carrying out a Principal Component Analysis to 
investigate the degree of organisation in this data.    

On the observational front work in progress by us includes enlarging the sample size as more SDSS and 
\hi data [e.g. from the Bonn 100 metre Multibeam Survey (Kerp et al. in preparation) becomes available. Our automated 
SDSS photometric pipeline for large galaxies should greatly accelerate progress (West et al. 2008). 
\hi interferometry is being obtained, first to sort out identifications in tight groups and second to measure 
more actual \hi radii to get a better understanding of the Minchin-Rosenberg effect. So far we have exploited neither 
the colours nor the fibre spectra but work is in hand to include them too (West et al. in preparation). Statistical 
work on mean cosmic properties, such as the mean \hi cosmic density,  depends on a clear understanding of ones 
selection effects and this is underway too (Garcia-Appadoo et al. in preparation). We are also working on the reverse List, 
that is to say a list of objects in the SDSS survey which we would have expected to contain measurable \hie, 
and thus show up in the Equatorial Survey, but which do not (West \& Garcia-Appadoo in preparation). 

\section*{Acknowledgments}

MJD would like to thank the UK Particle Physics and Astronomy Research Council (PPARC) for grants towards 
the support of both construction and exploitation of the HIPASS and HIJASS (HIPASS Northern equivalent at 
Jodrell Bank) surveys, and especially thank the HIPASS team, notably Professors Lister Staveley-Smith, 
Ron Ekers and Dr A. E. Wright of the Australia Telescope at CSIRO Radiophysics in Sydney, for the inspiration 
and drive which first got the Multibeam project going.

This publication makes use of data products from the Two Micron All Sky Survey, which is a joint project 
of the University of Massachusetts and the Infrared Processing and Analysis Center/California Institute 
of Technology, funded by the National Aeronautics and Space Administration and the National Science Foundation.

Funding for the SDSS and SDSS-II has been provided by the Alfred P. Sloan Foundation, the Participating 
Institutions, the National Science Foundation, the U.S. Department of Energy, the National Aeronautics 
and Space Administration, the Japanese Monbukagakusho, the Max Planck Society, and the Higher Education 
Funding Council for England. The SDSS Web Site is http://www.sdss.org/.

The SDSS is managed by the Astrophysical Research Consortium for the Participating Institutions. The 
Participating Institutions are the American Museum of Natural History, Astrophysical Institute Potsdam, 
University of Basel, University of Cambridge, Case Western Reserve University, University of Chicago, 
Drexel University, Fermilab, the Institute for Advanced Study, the Japan Participation Group, Johns 
Hopkins University, the Joint Institute for Nuclear Astrophysics, the Kavli Institute for Particle 
Astrophysics and Cosmology, the Korean Scientist Group, the Chinese Academy of Sciences (LAMOST), 
Los Alamos National Laboratory, the Max-Planck-Institute for Astronomy (MPIA), the Max-Planck-Institute 
for Astrophysics (MPA), New Mexico State University, Ohio State University, University of Pittsburgh, 
University of Portsmouth, Princeton University, the United States Naval Observatory, and the University 
of Washington.

\end{document}